\documentclass[10pt]{sig-alternate-05-2015}
\pdfoutput=1

\newif \ifdraft \drafttrue
\newif \ifappendix \appendixtrue

\usepackage{csquotes}
\usepackage{xspace}
\usepackage{ragged2e}
\usepackage{fixltx2e}
\usepackage{mdframed}
\usepackage{stmaryrd}
\usepackage{calc}
\usepackage[makeroom]{cancel}
\usepackage{algorithm}
\usepackage[noend]{algpseudocode}
\usepackage{makecell}
\usepackage{lineno}

\usepackage{enumitem}
\usepackage[justification=centering, font=small]{caption}
\usepackage[export]{adjustbox}
\usepackage{bm}
\usepackage{amstext}
\usepackage{tabularx}
\usepackage{balance}
\usepackage{authblk}
\usepackage{multirow}
\usepackage{subfig}
\usepackage[T1]{fontenc}
\usepackage{lmodern}
\usepackage{hyphenat}

\makeatletter
\renewcommand\AB@affilsepx{ , \protect\Affilfont}
\makeatother

\makeatletter
\renewcommand{\ALG@name}{SNAP-Policy}
\makeatother

\setlist[enumerate,1]{%
  label=\arabic*.,
}

\newlist{inlinelist}{enumerate*}{1}
\setlist*[inlinelist,1]{%
  label=(\roman*),
}

\newcommand{\lang}{SNAP\xspace}
\newcommand{\Lang}{SNAP\xspace}
\newcommand{\xFDD}{xFDD\xspace}
\newcommand{\xFDDs}{xFDDs\xspace}
\newcommand{\Tunnel}[1][footnotesize]{\codeb[#1]{DNS-tunnel-detect}\xspace}
\newcommand{\tsub}{\textsubscript}
\newcommand{\set}[1]{\{#1\}}
\newcommand{\sequence}[1]{<#1>}

\newcommand{\fddl}{\sqsubset}
\newcommand{\atomic}{\mathsf{atomic}}
\newcommand{\snaptitle}[1]{\noindent {\bf #1}}

\newcommand{\eval}{\mathsf{eval}}
\newcommand{\evale}{\mathsf{eval}_e}
\renewcommand{\merge}{\mathsf{merge}}
\newcommand{\elog}{\mathsf{E}}

\newcommand{\IfElse}[3]{\text{if } #1 \text{ then } #2 \text{ else } #3}

\newcommand{\ignore}[1]{}

\newcommand{\cam}[1]{#1}

\usepackage[hang,flushmargin]{footmisc} 
\usepackage{textcomp}
\newcommand{\cref}[1]{\textsection\ref{#1}}

\newsavebox\CBox
\newcommand\hcancel[2][0.5pt]{%
	\ifmmode\sbox\CBox{$#2$}\else\sbox\CBox{#2}\fi%
	\makebox[0pt][l]{\usebox\CBox}%
	\rule[0.5\ht\CBox-#1/2]{\wd\CBox}{#1}}

\newcommand{\codeb}[2][footnotesize]{\begin{#1}\texttt{#2}\end{#1}}
\newcommand{\match}[2]{#1 = #2}
\newcommand{\modify}[2]{#1 <- #2}

\newcommand{\union}[2]{#1+#2}
\newcommand{\seq}[2]{#1;#2}
\newcommand{\inters}[2]{#1 \& #2}
\newcommand{\id}{id}
\newcommand{\drop}{drop}

\newcommand{\boldifelse}[3]{{\textbf{if}} #1 \textbf{then} #2 \textbf{else} #3}

\newcommand{\pp}{\text{\codeb{++}}}
\newcommand{\mm}{\text{\codeb{-{}-}}}

\newcommand{\Expr}{\mathsf{Expr}}
\newcommand{\Pol}{\mathsf{Pol}}
\newcommand{\Pred}{\mathsf{Pred}}
\newcommand{\Packet}{\mathsf{Packet}}
\newcommand{\Log}{\mathsf{Log}}

\newcommand{\Store}{\mathsf{Store}}
\newcommand{\Val}{\mathsf{Val}}
\newcommand{\consistent}{\mathsf{consistent}}
\newcommand{\harpoon}{\overset{\rightharpoonup}}

\newcommand{\stgraph}{\textproc{st-dep}}
\newcommand{\R}{\textproc{r}}
\newcommand{\W}{\textproc{w}}

\newcommand{\highlighttext}[2][yellow]{{\setlength{\fboxsep}{0pt}\colorbox{#1}{#2}}}

\newcommand{\highlight}[2][yellow]{\mathchoice%
  {{\setlength{\fboxsep}{0pt}\colorbox{#1}{$\displaystyle#2$}}}%
  {{\setlength{\fboxsep}{0pt}\colorbox{#1}{$\textstyle#2$}}}%
  {{\setlength{\fboxsep}{0pt}\colorbox{#1}{$\scriptstyle#2$}}}%
  {{\setlength{\fboxsep}{0pt}\colorbox{#1}{$\scriptscriptstyle#2$}}}}

\newcommand{\footnotefont}{\fontsize{8}{10} \selectfont}
\let\oldfootnotesize\footnotesize
\renewcommand*{\footnotesize}{\oldfootnotesize \fontsize{9}{10}\selectfont}

\date{}
\title{\ttlfnt{SNAP: Stateful Network-Wide \\Abstractions for Packet Processing}}
\author[1]{Mina Tahmasbi Arashloo}
\author[1]{Yaron Koral} 
\author[2]{Michael Greenberg}
\author[1]{Jennifer Rexford}
\author[1]{David Walker}
\affil[1]{Princeton University}
\affil[2]{Pomona College}
\renewcommand\Affilfont{\small}
\begin{document}


\maketitle

\begin{sloppypar}

\begin{abstract}

Early programming languages for software-defined networking (SDN) were built
on top of the simple match-action paradigm offered by OpenFlow
1.0. However, emerging hardware and software switches offer much 
more sophisticated support for persistent state in the data plane, without involving a central
controller. Nevertheless, managing stateful, distributed systems
efficiently and correctly
is known to be one of the most challenging programming
problems. To simplify this new SDN problem, we introduce
SNAP.  

SNAP offers a simpler ``centralized'' stateful
programming model, by allowing programmers
to develop programs on top of \emph{one} big switch rather than \emph{many}.  
These programs may contain reads and writes to global, persistent arrays,
and as a result, programmers can implement a broad range of
applications, from stateful
firewalls to fine-grained traffic monitoring.
The SNAP compiler relieves programmers of having to worry
about how to distribute, place, and optimize access
to these stateful arrays by doing it all for them. 
More specifically, the compiler discovers read/write 
dependencies between arrays and translates one-big-switch programs into 
an efficient internal representation based on
a novel variant of binary decision diagrams. This internal representation 
is used to construct a mixed-integer linear program, which jointly
optimizes the placement of state and the routing of traffic
across the underlying physical topology.
We have implemented a prototype compiler and applied it to
about 20 SNAP programs over various topologies to demonstrate our
techniques' scalability.

\end{abstract}

\ccsdesc[500]{Networks~Programming interfaces}
\ccsdesc[500]{Networks~Programmable networks}
\ccsdesc[300]{Networks~Network control algorithms}
\ccsdesc[300]{Networks~In-network processing}
\ccsdesc[300]{Networks~Network management}

\printccsdesc
\keywords{\lang, Network Programming Language, Stateful Packet Processing, One Big Switch, Software Defined Networks, Optimization}

\newpage
\section{Introduction}\label{sec:intro}

The first generation of programming languages for software-defined networks (SDNs)~\cite{nox,frenetic,maple,pyretic,Kinetic} was built on top of OpenFlow 1.0, which offered simple match-action processing of packets.  As a result, these  systems were partitioned into (1) a stateless packet-processing part that could be analyzed statically, compiled, and installed on OpenFlow switches, and (2) a general stateful component that ran on the controller.  

This ``two-tiered'' programming model can support any network functionality by running the stateful portions of the program on the controller and modifying the stateless packet-processing rules accordingly.  However, simple stateful programs, such as detecting SYN floods or DNS amplification attacks, cannot
be implemented \emph{efficiently} because packets must go back-and-forth to the controller, incurring significant delay.  Thus, in practice, stateful controller programs are limited to those that do not require \emph{per-packet} stateful processing.

Today, however, SDN technology has advanced considerably: there is a raft of new proposals for switch interfaces that \emph{expose persistent state on the data plane}, including those in P4~\cite{p4}, OpenState~\cite{openstate}, POF~\cite{pof}, \cam{Domino~\cite{domino}}, and Open vSwitch~\cite{openvswitch}.
Stateful programmable data planes enable us to \emph{offload} programs that require \emph{per-packet} stateful processing 
onto switches, subsuming a variety of functionality normally relegated to middleboxes.
However, the mere existence of these stateful mechanisms does not make networks of these devices
easy to program. In fact, programming distributed collections of stateful devices is typically one of the
most difficult kinds of programming problems. We need new languages and abstractions to help us manage the
complexity and optimize resource utilization effectively.

For these reasons, we have developed \lang, a new language that allows programmers to mix 
primitive stateful operations with pure packet processing.  However, rather than ask programmers
to program a large, distributed collection of independent, stateful devices manually, we provide 
the abstraction that the network is \emph{one} big switch (OBS).
Programmers can allocate persistent arrays on that OBS
, and do not have to worry about where or how such arrays are stored in the 
physical network.
Such arrays can be indexed by fields in incoming packets and modified over time as network conditions  change. Moreover, if multiple arrays must be updated simultaneously, we provide  a form of \emph{network transaction} to ensure such updates occur atomically.  As a result,
it is easy to write \lang programs that learn about the network environment and record its state, 
store per-flow information or statistics, or implement a variety of stateful mechanisms.

While it simplifies programming, 
the OBS model, together with the stateful primitives,
generates implementation challenges.
In particular, multiple flows may depend upon the same \cam{state}.  
To process these flows correctly and efficiently, the compiler must simultaneously determine which flows
depend upon which components, how to route those flows, and where to place the 
components. 
Hence, to map OBS programs to concrete topologies, the \lang compiler  discovers read-write dependencies between statements. It then translates the program into an \xFDD, a variant of forwarding decision diagrams (FDDs)~\cite{FastNetKATCompiler} extended to incorporate stateful operations.  Next, the compiler generates a system of integer-linear equations that jointly optimizes array placement and traffic routing.  Finally, the compiler generates the switch-level configurations from the \xFDD and the optimization results.  We assume that the switches chosen for array placement 
support persistent programmable state; other switches 
can still play a role in routing flows efficiently through the state variables.
Our main contributions are:
\begin{itemize}[leftmargin=*]

\item A stateful and compositional SDN programming language with persistent 
global arrays, a one-big-switch programming model, and network transactions. (See \cref{sec:example}
for an overview and \cref{sec:language} for more technical details.)

\item Algorithms for compiling \lang programs into low-level switch mechanisms (\cref{sec:compilation}): (i) 
an algorithm for compiling \lang programs into an intermediate representation that detects
program errors, such as race conditions introduced by parallel access to stateful components, using 
 our extended forwarding decision diagrams (\xFDD) and (ii)
an algorithm to generate
a mixed integer-linear program, based on the \xFDD, 
which jointly decides array placement and routing while minimizing network congestion 
and satisfying the constraints necessary for network transactions. 

\item An implementation and evaluation of our language and compiler
using about 20 applications.
(\cref{sec:implementation}, \cref{sec:evaluation}).

\end{itemize}

We discuss various data-plane implementations for \lang, how \cam{\lang} relates to middleboxes, and possible extensions in \cref{sec:discussion}, discuss related work in 
\cref{sec:related}, and conclude in \cref{sec:conclusion}.
\section{SNAP System Overview}\label{sec:example}

\newenvironment{snappolicy}[1][htb]
{\renewcommand{\algorithmcfname}{SNAP-Policy}
	\begin{algorithm}[#1]%
	}{\end{algorithm}}

This section overviews the key concepts in our language 
and compilation process using example programs.

\subsection{Writing \lang Programs}
\label{sec:e2e_example}

\begin{figure}[t!]
\mdfsetup{
skipabove = 0cm,
skipbelow =  0.1mm,
innerrightmargin=-2mm}
\begin{mdframed}
\centering
\codeb[footnotesize]{\textbf{DNS-tunnel-detect}}\\
\end{mdframed}

\mdfsetup{
skipabove = 0cm,
skipbelow =  3mm,
innerrightmargin=-2mm}
\begin{mdframed}
\raggedright
\begin{internallinenumbers}
\setlength\linenumbersep{-0.5mm}
\setlength\leftskip{0.2cm}
\codeb[scriptsize]{
\boldifelse{\inters{\match{dstip}{10.0.6.0/24}}{\match{srcport}{53}}}
{\\ \hspace{0.3cm}
	\seq{\seq{\modify{orphan[dstip][dns.rdata]}{True}}\\ \hspace{0.3cm}
		{susp-client[dstip]++}}\\ \hspace{0.3cm}
	{\boldifelse{\match{susp-client[dstip]}{\emph{threshold}}}
		{\\ \hspace{0.8cm}\modify{blacklist[dstip]}{True}\\ \hspace{0.3cm}}
		{\id\\}
	}
	\hspace{-0.2cm}
}
{	\\ \hspace{0.3cm}
	\boldifelse{\inters{\match{srcip}{10.0.6.0/24}}
		{orphan[srcip][dstip]}\\\hspace{0.3cm}}
	{\seq{\modify{orphan[srcip][dstip]}{False}}{\\ \hspace{1cm}
		susp-client[srcip]\mm \\ \hspace{0.1cm}
		}}
		{\id\\}
}
}
\end{internallinenumbers}
\end{mdframed}
\caption{\lang implementation of \Tunnel.}
\label{fig:tunnel-code}
\end{figure}

\snaptitle{DNS tunnel detection.} The DNS protocol is designed to resolve information about domain names. 
\cam{Since it is not intended for general data transfer, DNS often
draws less attention in terms of security monitoring than other
protocols, and is used by attackers to bypass security policies and leak information. }
Detecting DNS tunnels is one of many
real-world scenarios that require
state to track the properties of network flows~\cite{chimera}.  
The following steps can be used to detect DNS tunneling~\cite{chimera}:
\begin{enumerate}
	\item For each client, keep track of the IP addresses resolved by DNS responses.
	\item For each DNS response, increment a counter.  This counter
              tracks the number of resolved IP addresses that a client does not use.
	\item When a client sends a packet to a resolved IP address, 
              decrement the counter for the client.
	\item Report tunneling for clients that exceed a threshold for resolved, but unused IP addresses.
\end{enumerate}

\begin{figure}[t!]
	\centering
	\includegraphics[width = 0.5\columnwidth]{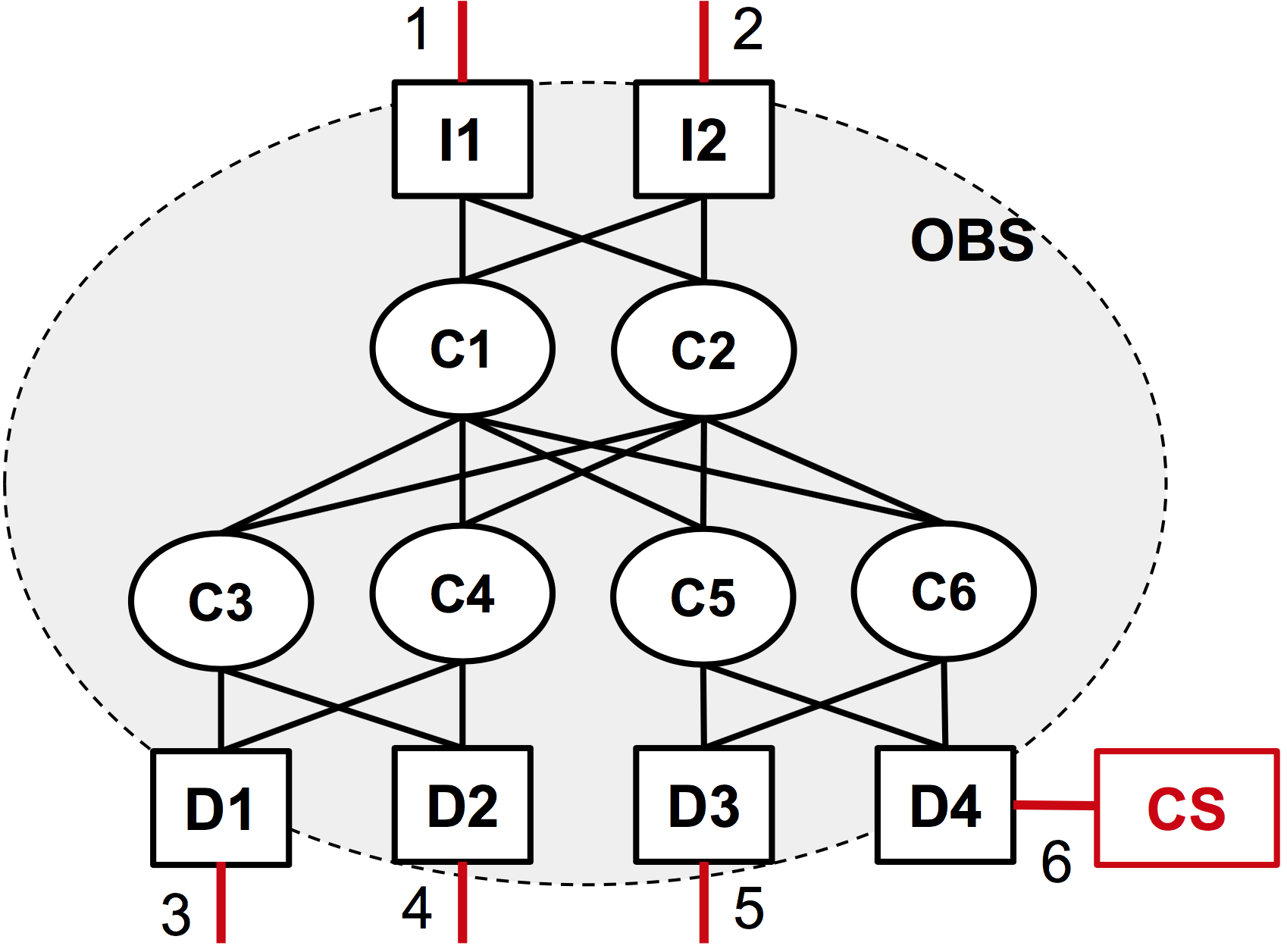}
	\caption{Topology for the running example.}
	\label{fig:example-topo}
\end{figure}

Figure~\ref{fig:tunnel-code} shows a \Lang implementation of the above steps
that detects DNS tunnels to/from the %
CS department subnet 10.0.6.0/24 (see Figure~\ref{fig:example-topo}).
Intuitively, a SNAP program can be thought of as 
a function that takes in a packet plus the current state of the
network and produces a set of transformed packets as well as updated state.
The incoming packet is read and written by referring to its fields 
(such as \codeb{dstip} and \codeb{dns.rdata}).
The ``state'' of the network is read and written by referring to user-defined, array-based, global variables
(such as \codeb{orphan} or \codeb{susp-client}).
Before explaining the program in detail, note that it does not refer to specific network
device(s) on which it is implemented.  \Lang programs are
expressed as if the network was \emph{one-big-switch} (OBS) connecting
edge ports directly to each other.
Our compiler automatically distributes the program across 
network devices, freeing programmers from such details and making SNAP programs portable across
topologies.

The \Tunnel program examines two kinds of packets: incoming DNS responses 
(which may lead to possible DNS tunnels) and outgoing packets to resolved IP addresses.
Line 1 checks whether the input packet is a DNS response
to the CS department. 
The condition in the \codeb{if} statement is an example
of a simple \emph{test}.  Such tests can involve any boolean combination of 
packet fields.\footnote{\footnotefont {The design of the language is unaffected by the chosen set of fields.
For the purposes of this paper, we assume a rich set of fields, e.g. DNS response data.
New architectures such as P4~\cite{p4} have programmable parsers that allow users to customize
their applications to the set of fields required.}}
If the test succeeds, the packet could potentially belong to a DNS tunnel,
and will go through the detection steps (Lines 2--6).
Lines 2--6 use three global variables to keep track of DNS queries. Each variable 
is a mapping between keys and values, persistent across multiple packets. 
The \codeb{orphan} variable, for example,
maps each pair of IP addresses to a boolean value.  If \codeb{orphan[c][s]} is
\codeb{True} then \codeb{c} has received a DNS
response for IP address \codeb{s}. The variable \codeb{susp-client}
maps the client's IP to the number of 
DNS responses it has received
but not accessed yet.
If the packet is not a DNS response, a different test is performed, 
which includes a stateful test over \codeb{orphan} (Lines 8).
If the test succeeds, 
the program updates \codeb{orphan[srcip][dstip]} to \codeb{False} and 
decrements \codeb{susp-client[srcip]} (Lines 10--11).
This step changes the global state and thus, affects the processing of future packets. 
Otherwise, the packet is left unmodified --- \codeb{id} (Line 12) is a no-op.

\snaptitle{Routing.}  \Tunnel  
cannot stand on its own---it does not explain where to forward  packets.
In \Lang, we can easily \textit{compose} it with a  forwarding policy.
Suppose our target network is the simplified campus topology depicted in 
Figure~\ref{fig:example-topo}.  Here, $I_1$ and $I_2$ are connections to the Internet, 
and $D_1$--$D_4$ represent edge switches in the departments, with $D_4$ connected to the CS building.
$C_1$--$C_6$ are core routers connecting the edges. 
External ports (marked in red) are numbered 1--6
and IP subnet \codeb{10.0.i.0/24} is attached to port \codeb{i}.
The \codeb{assign-egress} program assigns outports to packets based on their destination IP
address:
{
\mdfsetup{
skipabove=4mm,
skipbelow=4mm,
rightmargin=.35cm,
leftmargin=.35cm,
}
\begin{mdframed}
\raggedright
\codeb[scriptsize]{
\hspace{-0.3cm} assign-egress = \boldifelse{\match{dstip}{10.0.1.0/24}\\}
   	{\modify{outport}{1}\\}
	{\boldifelse{\match{dstip}{10.0.2.0/24}}
		  {\modify{outport}{2}\\}
		  {...}\\
 	}
	\textbf{else }\boldifelse{\match{dstip}{10.0.6.0/24}}
					 {\modify{outport}{6} \\}
					 {\drop \\}
} 
\end{mdframed}
}

Note that the policy is independent of the internal network structure,
and recompilation is needed only if
the topology changes. 
By combining \Tunnel with \codeb{assign-egress}, we have implemented a useful
end-to-end program:
\codeb[footnotesize]{
\seq{\Tunnel[footnotesize]}{assign-egress}}.

\snaptitle{Monitoring.}  Suppose the operator wants to monitor packets
entering the network at each ingress port (ports 1-6).  She 
might use an array indexed by \codeb{inport} and increment the corresponding element
on packet arrival:  \codeb[footnotesize]{count[inport]++}.  
Monitoring should take place \emph{alongside} the rest of the program; thus, she might combine it using parallel
composition (\codeb[footnotesize]{+}): \codeb[footnotesize]{(\Tunnel[footnotesize] + } \codeb[footnotesize]{count[inport]++);} \codeb[footnotesize]{assign-egress}.
Intuitively, \codeb[footnotesize]{p + q} makes a copy of the incoming
packet and executes both \codeb{p} and \codeb{q} on it simultaneously.  

Note that it is not always legal to compose two programs in parallel.
For instance, if one writes to the same global variable that the other reads,
there is a race condition, which leads to ambiguous state in the final program.
Our compiler detects such race conditions and rejects ambiguous programs.

\snaptitle{Network Transactions.}
Suppose that an operator sets up a
honeypot at port 3 with IP subnet 10.0.3.0/25.  The following program records,
per inport, the IP and dstport of the last packet destined to the honeypot: \\

{
\mdfsetup{
skipabove=0mm,
skipbelow=5mm,
rightmargin=1cm,
leftmargin=1cm,
}
\begin{mdframed}
\raggedright
\codeb[scriptsize]{
\hspace{-2mm}\boldifelse{\match{dstip}{10.0.3.0/25} \\}
	        {\seq{\modify{hon-ip[inport]}{srcip}}{ \\ \hspace{0.6cm} \modify{hon-dstport[inport]}{dstport}} \\}
	        {\id} \\
}
\end{mdframed} 
}

Since this program processes many packets simultaneously, 
it has an implicit race condition:  if packets $p_1$ and $p_2$, both
destined to the honeypot, enter the network from port 1 and get reordered,
each may visit \codeb{hon-ip} and \codeb{hon-dstport}
in a different order (if the variables reside in different locations). Therefore, it is possible that \codeb{hon-ip[1]} contains the 
source IP of $p_1$ and \codeb{hon-dstport[1]}
the destination \cam{port} of $p_2$ while the operator's intention was 
that both variables refer to the same packet.
To establish such properties for a
collection of state variables, programmers can use \emph{network transactions} by
simply enclosing a series of statements in an \codeb{atomic} block. Atomic
blocks  co-locate their enclosed state variables so that a series of updates can be made to appear atomic.

\subsection{Realizing Programs on the Data Plane}
\label{subsec:compiler-overview}
Consider
\codeb[footnotesize]{\Tunnel[footnotesize];}
\codeb[footnotesize]{assign-egress}.
To distribute this program across network
devices, the \Lang compiler should
decide (i) where to place state variables (\codeb{orphan}, \codeb{susp-client},
and \codeb{blacklist}), and (ii) how packets should be routed across the
physical network. 
These decisions should be made in such a way 
that each packet passes through
devices storing \emph{every} state variable it \emph{needs}, 
in the correct \emph{order}. 
Therefore, the compiler needs information about which packets
need which state variables.
In our example program, for instance, packets with
\codeb{\match{dstip}{10.0.6.0/24}} and \codeb{\match{srcport}{53}}
need to pass all three state variables,
with \codeb{blacklist} accessed after the other two.  

\begin{figure}[t!]
\centering
\includegraphics[width = 0.78\columnwidth]{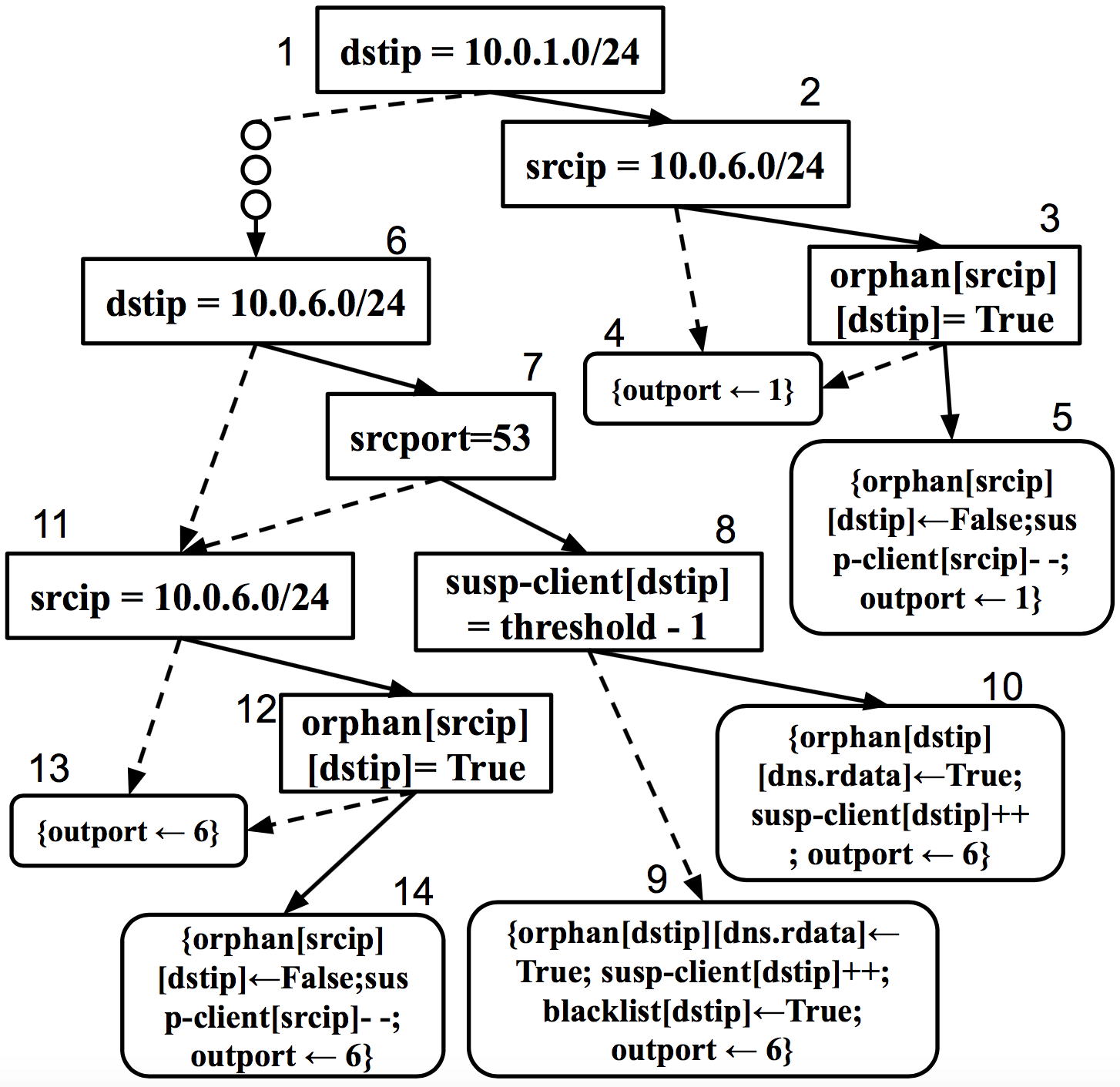}
\caption{The equivalent \xFDD for \\ \codeb[scriptsize]{\Tunnel[scriptsize]; assign-egress}
}
\label{fig:ex-fdd3}
\end{figure}

\snaptitle{Program analysis.} To extract the above information, we transform the program to an
intermediate representation called \emph{extended forwarding} \emph{decision diagram (xFDD)} 
(see Figure~\ref{fig:ex-fdd3}). 
FDDs were originally introduced in an earlier
work~\cite{FastNetKATCompiler}.  We extended FDDs
in \lang to support stateful packet processing.
An xFDD is like a binary decision diagram: each intermediate node is a
test on either packet fields or state variables.
The leaf nodes are sets of action sequences, rather than merely `true'
and `false' as in a BDD~\cite{bdds}.  Each interior node has two successors:
\emph{true} (solid line), which determines the rest of the
forwarding decision process for inputs passing the test, and
\emph{false} (dashed line) for failed cases.
xFDDs are constructed compositionally; the xFDDs for different parts of
the program are combined to construct the final xFDD. 
Composition is particularly more involved with
stateful operations: the same state variable may be referenced in two xFDDs
with different header fields, e.g., once as \codeb{s[srcip]} and then
as \codeb{s[dstip]}. How can we know whether or not those fields are
equal in the packet?
We add a new kind of test,
over pairs of packet fields (\codeb{srcip = dstip}), and new
ordering requirements on the xFDD structure.

Once the program is transformed to an \xFDD, 
we analyze the \xFDD to extract information about 
which groups of packets need which state variables.
In Figure~\ref{fig:ex-fdd3}, for example, leaf number 10 is on the true branch of
\codeb{dstip=10.0.6.0/24} and \codeb{srcport=53}, which indicates that all packets
with this property may end up there.  These packets need 
\codeb{orphan}, because it is modified, and \codeb{susp-client}, because it 
is both tested and modified on the path. 
We can also deduce these packets can enter the network
from any port and the ones that are not dropped will exit port 6.
Thus, we can \cam{use the \xFDD to figure out which packets
need which state variables,
aggregate this information across OBS
ports,} 
and choose paths for traffic between these ports accordingly.

\snaptitle{Joint placement and routing.}
At this stage, the compiler has the information it needs to
distribute the program. It uses a mixed-integer linear
program (MILP) that solves an extension of the multi-commodity flow
problem to \emph{jointly} decide state placement and routing while
minimizing network congestion.
The constraints in the MILP guarantee that
the selected paths for each pair of OBS ports take 
corresponding packets through 
devices storing every state variable that they need, 
in the correct order. Note that the \xFDD analysis 
can identify cases in which both directions of a
connection need the same state variable $s$, so the MILP
ensures they both traverse the device holding $s$.

In our example program, 
the MILP places all state variables on D\tsub{4},
which is the optimal location as all packets to and from the protected 
subnet must flow through D\tsub{4}.\footnote{
\footnotefont
State can be spread out across the network.
It just happens that in this case, one location turns out to be optimal.
} Note that this is not obvious
from the \Tunnel code alone, but rather from its \emph{combination} with 
\codeb{assign-egress}.
This highlights the fact that in \Lang, program components can be written in a modular way, 
while the compiler makes globally optimal decisions using information from all parts.
The optimizer also decides forwarding paths between
external ports. For instance, traffic from
$I_1$ and $D_1$ will go through $C_1$ and $C_5$ to reach $D_4$. The path from $I_2$ and $D_2$
to $D_4$ goes through $C_2$ and $C_6$, and $D_3$ uses $C_5$ to reach $D_4$. The paths between
the rest of the ports are also determined by the MILP in a way that minimizes 
link utilization. The compiler takes state placement and routing results from the 
MILP, partitions the program's intermediate representation (\xFDD) among switches,
and generates rules for the controller to push to all stateless and stateful switches in the 
network. 

\snaptitle{Reacting to network events.}
The above phases only run if the operator 
changes the OBS program.
Once the program compiles, and to respond to network events 
such as failures or traffic shifts, we use
a simpler and much faster version of the MILP that given
the current state placement, only re-optimizes for routing. 
Moreover, with state on the data plane, policy changes become considerably
\emph{less frequent} because the policy, and consequently switch configurations,
\emph{do not} change upon changes to state.
In \Tunnel[footnotesize], for instance,
attack detection and mitigation are both captured in the program itself, happen
\emph{on the data plane}, and therefore react rapidly to malicious activities in the
network. 
This is in contrast to the case where all the state is on the 
controller. There, the policy needs to change and recompile multiple times
both during detection and on mitigation, to reflect the state changes
on the controller in the rules on the data plane.
\section{\Lang}\label{sec:language}

\Lang is a high-level language with two key features: programs
are \emph{stateful} and are written in terms of an abstract network
topology comprising a \emph{one-big-switch} (OBS). It has an algebraic
structure patterned on the NetCore/NetKAT family of
languages~\cite{NetCore,netkat},
with each program comprising one or more \emph{predicates} and
\emph{policies}.
\Lang's syntax is in Figure~\ref{fig:syntax}. Its semantics is 
defined through an evaluation function ``$\eval{}$.'' $\eval{}$ determines,
in mathematical notation, how an input packet should be processed
by a \Lang program. Note that this is part of the \emph{specification} of the language,
\emph{not} the implementation. Any implementation of \Lang, including ours, 
should ensure that packets are processed as defined by the 
$\eval{}$ function: when we talk about 
``running'' a program on a packet, we mean calling $\eval{}$ on
that program and packet.
We discuss $\eval{}$'s most interesting cases here; see appendix~\ref{app:semantics} for a full definition.

$\eval{}$ takes the \Lang term of interest,
a packet, and a starting 
state and yields a set of packets and an output state. 
To properly define the semantics of multiple updates 
to state when programs are composed, we need to know
the reads and writes to state variables performed by each program while 
evaluating the packet. Thus, $\eval{}$ also returns a \emph{log}
containing this information.  
It adds  ``$R\,s$'' to the log  whenever a read from
state variable $s$ occurs,
and ``$W\,s$'' on writes.
Note that these logs are part of our formalism, but not our
implementation.
We express the program state as a dictionary that maps state
variables to their contents.  The contents of each state variable is
itself a mapping from values to values.
Values are defined as
packet-related fields (IP address, TCP ports, MAC addresses, 
DNS domains) along with integers, booleans and 
vectors of such values. 

\begin{figure}
\scriptsize
\mdfsetup{
innertopmargin = -1mm,
innerleftmargin = 1mm,
skipbelow = 2mm,
}
\begin{mdframed}
\[\begin{array}{rcll}
  e \in \Expr & ::= &  v \, | \, f \, | \, \harpoon e & \\ 
  x, y \in \Pred & ::= & \id & \text{Identity} \\
  & | & \drop      & \text{Drop} \\ 
  & | & f = v  & \text{Test} \\ 
  & | & \neg x & \text{Negation} \\ 
  & | & x | y  & \text{Disjunction} \\ 
  & | & y \& x  & \text{Conjunction} \\
  & | & \highlight{s[e] = e} & \textbf{State Test} \\
  p, q \in \Pol & ::= & x & \text{Filter} \\
  & | & f \leftarrow v & \text{Modification} \\
  & | & p + q          & \text{Parallel comp.} \\ 
  & | & p ; q          & \text{Sequential comp.} \\ 
  & | & \highlight{s[e] \gets e}     & \textbf{State Modification} \\ 
  & | & \highlight{s[e]\codeb{++}} 		& \textbf{Increment value} \\
  & | & \highlight{s[e]\mm}			& \textbf{Decrement value} \\
  & | & \highlight{\IfElse{a}{p}{q}} & \textbf{Conditional} \\
  & | & \highlight{\atomic(p)}       & \textbf{Atomic blocks}
\end{array}\]
\end{mdframed}
\caption{\Lang's syntax. \highlighttext{Highlighted} items are not in
  NetCore.}
\label{fig:syntax}
\end{figure}

\snaptitle{Predicates.}
Predicates have a constrained semantics: they never update the
state (but may read from it), and either return the empty
set or the singleton set containing the input packet. That is,
they either pass or drop the input packet.
$\id$ passes the packet and
$\drop$ drops it.
The test $f=v$ passes a packet $pkt$ if the field
$f$ of $pkt$ is $v$.
These predicates yield empty logs.

The novel predicate in \Lang is the \emph{state test}, written $s[e_1]
= e_2$ and read ``state variable (array) $s$ at index $e_1$ equals $e_2$''.
Here $e_1$ and $e_2$ are \emph{expressions}, where an expression is
either a value $v$ (like an IP address or TCP port), a field $f$,
or a vector of them $\harpoon e$.
For $s[e_1] = e_2$, function $\eval$ evaluates $e_1$ and $e_2$ on
the input packet 
to yield two values
$v_1$ and $v_2$. 
The packet can pass if state variable $s$
indexed at $v_1$ is equal to $v_2$, 
and is dropped otherwise.
The returned log will include $R\,s$,
to record that the predicate read from the state variable $s$.

We evaluate negation $\neg x$ by running $\eval{}$ on $x$
and then
complementing the result, 
propagating whatever log $x$ produces. 
$x | y$ (disjunction) unions the
results of running $x$ and $y$ individually, doing the reads
of both $x$ and $y$.
$x\&y$ (conjunction) intersects the results of 
running $x$ and $y$ while doing the reads of $x$ and then $y$.

\snaptitle{Policies.}
Policies can modify packets and the state. Every predicate is a
policy---it simply makes no modifications.
Field modification $f \leftarrow v$ 
takes an input packet $pkt$ and yields a new packet,
$pkt'$, such that $pkt'.f = v$ but otherwise $pkt'$ is the same as
$pkt$. State update $s[e_1] \gets e_2$ passes the input packet through while (i)
updating the state so that $s$ at $\eval(e_1)$ is
set to $\eval(e_2)$, and (ii) adding $W\,s$ to the log. The $s[e]\codeb{++}$ (resp. $\mm$) operators increment (decrement) the value of $s[e]$ 
and add $W\,s$ to the log. 

Parallel composition $p + q$ runs $p$ and $q$ in parallel and tries to
merge the results. If the logs indicate a state read/write or
write/write conflict for $p$ and $q$ then there is
no consistent semantics we can provide, and we leave the semantics
undefined.
Take for example $(s[0] \gets 1) + (s'[0] \gets 2)$. 
There is no conflict if $s \neq s'$.
However, 
the state updates conflict if $s = s'$. 
There is no good choice here, so we leave
the semantics undefined and raise \emph{compile error} in the implementation.

Sequential composition $p;q$ runs $p$ and then runs $q$ on each packet
that $p$ returned, merging the final results. We must 
ensure the runs of $q$ are pairwise consistent, or else we will have a
read/write or write/write conflict. 
For example, let $p$ be $(f \leftarrow 1 + f \leftarrow 2)$, 
and $pkt[f \mapsto v]$ denote ``update $pkt$'s $f$ field to $v$''.
Given a packet $pkt$, the policy $p$
produces two packets: $pkt_1 = pkt[f \mapsto 1]$ and $pkt_2 = pkt[f
  \mapsto 2]$.
Let $q$ be $s[0] \gets f$,
running $p;q$ fails because 
running $q$ on $pkt_1$ and $pkt_2$ updates $s[0]$ differently.
However, $p;q$ runs fine for $q = g \leftarrow 3$.

We have an explicit conditional ``$\IfElse{a}{p}{q}$,'' which indicates \emph{either} $p$ or $q$ are executed. Hence, both $p$ and $q$ can perform reads and writes to the same state.
We have a notation for \emph{atomic blocks}, written
$\atomic(p)$. As described in \cref{sec:example}, there is
a risk of inconsistency between state variables residing on 
different switches on a real network 
when many packets are in flight concurrently.
When compiling $\atomic(p)$, our compiler ensures that
all the state in $p$ is updated atomically (\cref{sec:compilation}).
\section{Compilation}\label{sec:compilation}

\begin{figure}[t!]
  \centering
  \includegraphics[width = .75\columnwidth]{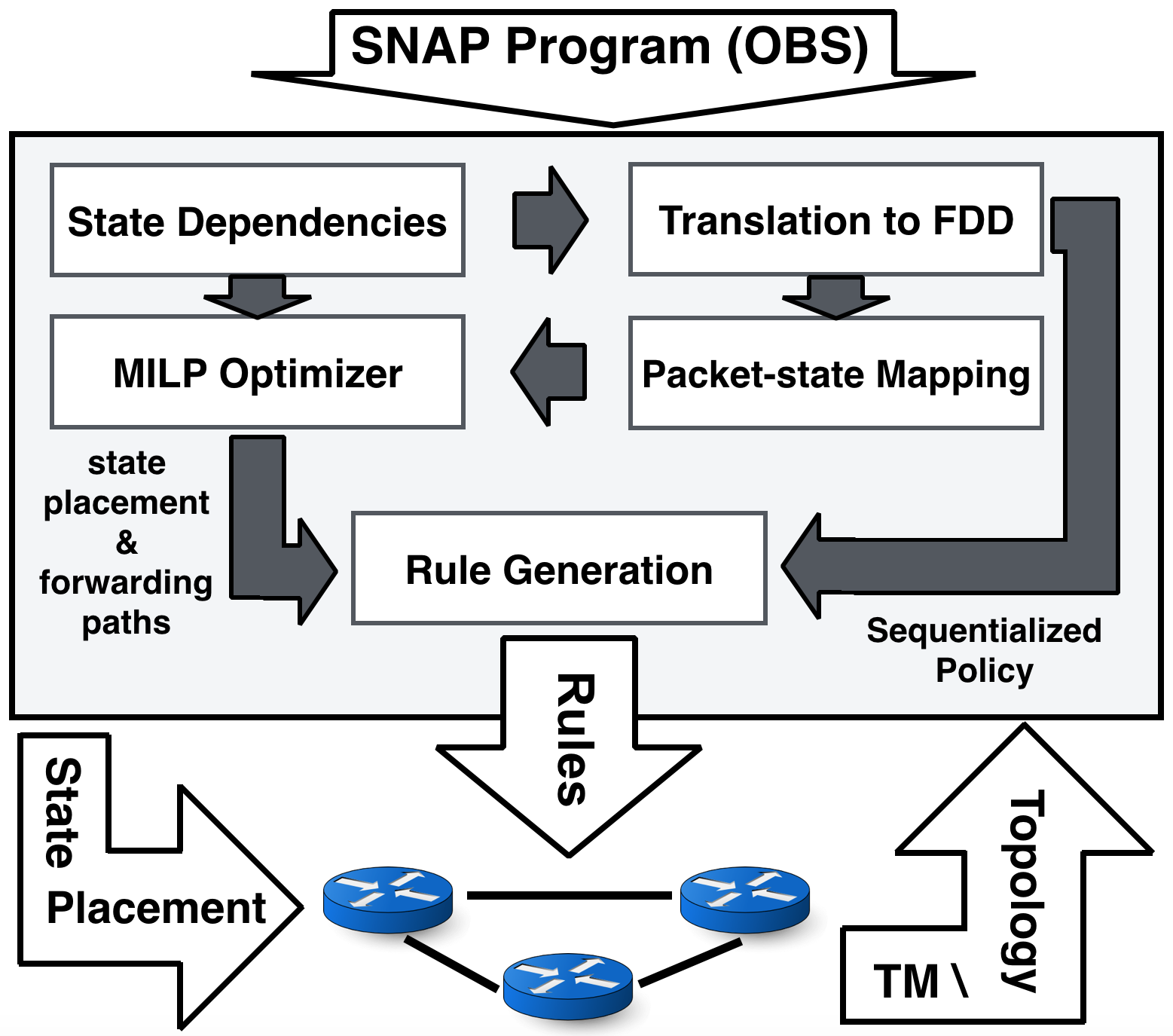}
  \caption{Overview of the compiler phases.}
  \label{fig:compilation}
\end{figure}

To implement a \Lang program specified on one big switch,
we must fill in two critical details: \emph{traffic
  routing} and \emph{state placement}.  The physical topology may
offer many paths between edge ports, and many possible locations for
placing state.\footnote{\footnotefont In this work, we assume each state
  variable resides in one place, though it is conceivable
  to distribute it (see \cref{sec:milp} and \cref{subsec:extensions}).}
The routing and placement problems interact: if two flows (with
different input and output OBS ports) both need some state variable
$s$, we should select routes for the two flows such that they pass
through a common location where we place $s$.
Further complicating the situation, the OBS program may specify that
certain flows read/write multiple state variables in a particular
order. The routing and placement on the physical topology must respect
that order.  In \Tunnel, for instance, routing must ensure that
packets reach wherever \codeb{orphan} is placed before
\codeb{susp-client}.
In some cases, two different flows may depend on
the same state variables, but in different orders.

We have designed a \emph{compiler} that translates OBS programs
into forwarding rules and state placements for a given topology.
As shown in Figure~\ref{fig:compilation},
the two key phases are (i) translation to \emph{extended forwarding decision
  diagrams} (xFDDs)---used as the intermediate representation of the program and to
  calculate which flows need which state
variables---and (ii) optimization via \emph{mixed integer linear program}
(MILP)---used to decide routing and state placement.
In the rest of this section, we present the compilation process in
phases, first discussing the analysis of state dependencies, followed
by the translation to xFDDs and the packet-state mapping, then the
optimization problems, and finally the generation of rules sent to the
switches.

\subsection{State Dependency Analysis}
\label{subsec:statedep}

Given a program, the compiler first performs \emph{state
  dependency analysis} to determine the ordering constraints on its state variables.
A state variable $t$ \emph{depends} on a state variable $s$ if the program writes to $t$
after reading from $s$. Any realization of the program on a
concrete network must ensure that $t$ does not come before $s$.  
Parallel composition, $p + q$, introduces no dependencies: if $p$
reads or writes state, then $q$ can run independently of that. 
Sequential composition $p;q$, on the other hand, introduces
dependencies: whatever reads are in $p$ must happen before writes
in $q$.
In explicit conditionals ``$\IfElse{a}{p}{q}$'', the writes in $p$ and $q$
depend on the condition $a$. 
Finally, atomic sections $\atomic(p)$ say that all state in $p$ is
inter-dependent. In \Tunnel, for instance,
\codeb{blacklist} is dependent on \codeb{susp-client}, itself
dependent on \codeb{orphan}.
This information is encoded as a dependency graph on 
state variables and is used to order the \xFDD structure (\cref{sec:fdds}),
and in the MILP (\cref{sec:milp}) to drive state placement.

\subsection{Extended Forwarding Decision Diagrams}
\label{sec:fdds}

\newcommand{\fdd}{\textproc{to-xfdd}}

\begin{figure}[t!]
\scriptsize
\mdfsetup{
innerleftmargin=0cm,
innertopmargin= -1mm,
skipbelow=-3mm,
rightmargin=0mm,
leftmargin=0mm,
}
\begin{mdframed}
\[
\begin{array}{rclr}
  d &::=& t~?~d_1 : d_2 \,|\, \set{as_1, \dots, as_n} & \text{xFDDs} \\
  t &::=& f = v \,|\, f_1 = f_2 \,|\, s[e_1] = e_2 & \text{tests} \\
  as & ::= & a \,|\, a;a & \text{action sequences} \\ 
  a &::=& \id \,|\, \drop \,|\, f \leftarrow v \,|\, s[e_1] \gets e_2 & \text{actions} \\
  & &  \,|\, s[e_1]\pp \,|\, s[e_1]\mm \\
\end{array}\]
\end{mdframed}
\mdfsetup{
rightmargin=0mm,
innertopmargin= -2mm,
skipabove=0mm,
skipbelow=2mm,
}
\begin{mdframed}
\[ 
\arraycolsep=1.2pt
\begin{array}{rcl}
\fdd(a) &=& \set{ a } \\
\fdd(f = v) &=& f = v ~?~ \set{\id} ~:~ \set{\drop} \\
\fdd(\neg x) &=& \ominus \fdd(x) \\
\fdd(s[e_1] = e_2) &=& s[e_1] = e_2 ~?~ \set{\id} ~:~ \set{\drop} \\
\fdd(\atomic(p)) &=& \fdd(p) \\
\fdd(p + q) &=& \fdd(p) \oplus \fdd(q) \\
\fdd(p;q) &=& \fdd(p) \odot \fdd(q) \\
\fdd(\IfElse{x}{p}{q}) 
  &=& (\fdd(x) \odot \fdd(p))  \\
   &\oplus {}& (\ominus \fdd(x) \odot \fdd(q))
\end{array} \]
\end{mdframed}
\caption{\xFDD syntax and translation.}
\label{fig:fdd-syntax}
\end{figure}

\begin{figure*}[t!]
\scriptsize

\begin{tabularx}{\textwidth}{|c|X|}
\hline

\arraycolsep=0.7pt
$\begin{array}{rcl}
\set{as_{11}, \cdots, as_{1n}} \oplus \set{as_{21}, \cdots, as_{2m}} & = & \set{as_{11}, \cdots, as_{1n}} \cup \set{as_{21}, \cdots, as_{2m}} \\
(t~?~d_1 : d_2) \oplus \set{as_1, \cdots, as_n} & = & (t~?~ d_1 \oplus \set{as_1, \cdots, as_n} : d_2 \oplus \set{as_1, \cdots, as_n}) \\\\
(t_1~?~ d_{11} : d_{12}) \oplus (t_2~?~d_{21} : d_{22}) & = &
\begin{cases}
(t_1 ~?~ d_{11} \oplus d_{21} : d_{12} \oplus d_{22}) & t_1 = t_2 \\
(t_1 ~?~ d_{11} \oplus (t_2~?~d_{21}:d_{22}) : d_{12} \oplus (t_2~?~d_{21}:d_{22}) & t_1 \fddl t_2 \\
(t_2 ~?~ d_{21} \oplus (t_1~?~d_{11}:d_{12}) : d_{22} \oplus (t_1~?~d_{11}:d_{12}) & t_2 \fddl t_1
\end{cases}
\end{array}$
& 
\arraycolsep=0.7pt
$\begin{array}{rcl}
\ominus \set{id} &= & \set{\drop} \\
\ominus \set{\drop} & = & \set{id}\\
\ominus (t?d_1:d_2) & = & (t?\ominus d_1:\ominus d_2)
\end{array}$
  \\ \hline
\end{tabularx}
\begin{tabularx}{\textwidth}{|XcX|c|}
\hline
&
\arraycolsep=3pt
$\begin{array}{lcl}
as \odot \set{as_1, \cdots, as_n} & = & \set{as \odot as_1, \cdots, as \odot as_n}\\
as \odot (t~?~d_1:d_2) & = & \text{(see explanations in \cref{sec:fdds})} \\
\set{as_1, \cdots, as_n} \odot d & = & (as_1 \odot d) \oplus \cdots \oplus (as_n \odot d) \\
(t~?~d_1:d_2) \odot d & = & (d_1 \odot d)|_t \oplus (d_2 \odot d)|_{\sim t} \\
\end{array}$ 
& &
\arraycolsep=1pt
$\begin{array}{rcl}
\set{as_1, \cdots, as_n}|_t & = & (t~?~\set{as_1, \cdots, as_n} : \set{\drop})\\\\
(t_1~?~d_1:d_2)|_{t_2} & =  & 
\begin{cases}
(t_1~?~d_1:\set{\drop}) & t_1 = t_2 \\
(t_2~?~(t_1~?~d_1:d_2):\set{\drop}) & t_2 \fddl t_1 \\
(t_1~?~d_1|_{t_2}:d_2|_{t_2}) & t_1 \fddl t_2 
\end{cases}
\end{array}$
\\
\hline 
\end{tabularx}
\caption{Definitions of \xFDD composition operators.}
\label{fig:fdd-normalization}
\end{figure*}

\begin{figure*}
\scriptsize
\newcommand{\refine}{\textproc{refine}}
\newcommand{\kw}[1]{\text{\textbf{#1}}}
\mdfsetup{
skipbelow = 0.05mm,
}
\begin{mdframed}
$ \begin{array}{rcl}
\oplus(\set{as_{11}, \cdots, as_{1n}}, \set{as_{21}, \cdots, as_{2m}}, context) & = & \set{as_{11}, \cdots, as_{1n}} \cup \set{as_{21}, \cdots, as_{2m}} \\
\oplus((t~?~d_1 : d_2), \set{as_1, \cdots, as_n}, context) & = & \kw{let } c_T = context.add(t) \kw{ in }\\
& & \kw{let } brch_T = \oplus (d_1, \set{as_1, \cdots, as_n}, c_T) \kw{ in }\\ 
& & \kw{let } c_F = context.add(\neg t) \kw{ in }\\
& & \kw{let } brch_F = \oplus (d_2, \set{as_1, \cdots, as_n}, c_T) \kw{ in }\\ 
& & (t~?~ brch_T : brch_F) \\\\
\oplus(d_1, d_2, context) & = & \kw{let } (t_1~?~ d_{11} : d_{12}) = \refine(d_1, context) \kw{ in }\\
& & \kw{let } (t_2~?~d_{21} : d_{22}) = \refine(d_2, context) \kw{ in } \\
& & \kw{let } c_T = \kw{ if } t_1 \fddl t_2 \kw{ then } context.add(t_1) \kw{ else } context.add(t_2) \kw{ in }\\
& & \kw{let } c_F = \kw{ if } t_1 \fddl t_2 \kw{ then } context.add(\neg t_1) \kw{ else } context.add(\neg t_2) \kw{ in }\\
& & \begin{cases}
(t_1 ~?~ \oplus (d_{11}, d_{21}, c_T) : \oplus(d_{12}, d_{22}, c_F)) & t_1 = t_2 \\
(t_1 ~?~  \oplus(d_{11}, (t_2~?~d_{21}:d_{22}), c_T) : \oplus(d_{12}, (t_2~?~d_{21}:d_{22}), c_F) & t_1 \fddl t_2 \\
(t_2 ~?~ \oplus(d_{21}, (t_1~?~d_{11}:d_{12}), c_T) : \oplus(d_{22}, (t_1~?~d_{11}:d_{12}), c_F) & t_2 \fddl t_1
\end{cases}
\end{array} $
\end{mdframed}
\mdfsetup{
skipabove = 0mm,
skipbelow = 3mm,
}
\begin{mdframed}
\[\begin{array}{rcl}
\refine(\set{as_1, \cdots, as_n}, context) & = & \set{as_1, \cdots, as_n} \\
\refine((t ~ ? ~ d_1 : d_2), context) & = & \kw{if } context.imply(t) \kw{ then } \refine(d_1, context) \\
& & \kw{else if } context.imply(\neg t) \kw{ then } \refine(d_2, context) \\
& & \kw{else } (t ~ ? ~ d_1 : d_2) \\
\end{array} \]
\end{mdframed}
\caption{A closer look at $\oplus$.}
\label{fig:fdd-normalization-detailed}
\end{figure*}

\cam{The input to the compiler is a \lang program, which can be a composition of several smaller
programs. The output, on the other end, is the distribution of the original policy across the network. 
Thus, in between, we need an intermediate representation for \lang programs that is both composable
and easily partitioned. This intermediate representation can help the compiler compose 
small program pieces into a unified representation, which can further be partitioned to get distributed
across the network. Extended forwarding decision diagrams (\xFDDs), which are introduced in this section,
are what we use as our internal representation of \lang programs and have both desired properties. 
They also simplify analysis of \lang programs for extracting packet-state mapping, which we discuss in~\cref{sec:psm} } 

Formally (see Figure~\ref{fig:fdd-syntax}), an xFDD is either a \emph{branch} $(t ~?~ d_1 : d_2)$, where $t$
is a test and $d_1$ and $d_2$ are xFDDs, or a \emph{set} of action sequences
$\set{as_1, \dots, as_n}$.
Each branch can be thought of as a conditional: if the test $t$ holds
on a given packet $pkt$, then the xFDD continues processing $pkt$ using
$d_1$; if not, 
processes $pkt$ using $d_2$.
There are three kinds of tests. 
The
\emph{field-value test} $f=v$ holds when $pkt.f$ is
equal to $v$.
The \emph{field-field test} $f_1=f_2$ holds when the
values in $pkt.f_1$ and $pkt.f_2$ are equal.
Finally, the \emph{state test} $s[e_1] = e_2$ holds
when the state variable $s$ at index $e_1$ is equal to $e_2$.
The last two tests are our extensions to FDDs. The state tests support
our stateful primitives, and as we show later in this section, the field-field
tests are required for correct compilation. 
Each leaf in an xFDD is a set of action sequences, with each action being
either the identity,  drop, field-update 
$f \leftarrow v$, or  state update $s[e_1] \gets e_2$, which is another extension to the original FDD.  

A key property of \xFDDs is that the order of their tests ($\fddl$)
must be defined in advance.  
\cam{This ordering is necessary to ensure that each test is present at most once on any
path in the final tree when merging two \xFDDs into one. Thus, \xFDD composition can be done
efficiently without creating redundant tests.}
In our \xFDDs, we ensure
that all field-value tests precede
all field-field tests, themselves preceding all state tests. 
Field-value tests themselves are ordered by fixing an arbitrary order on 
fields and values. Field-field tests are ordered similarly.
For state tests,
we first define a total order on state variables by looking at the dependency 
graph from \cref{subsec:statedep}. We break \cam{the dependency graph} into strongly
connected components (SCCs) and fix an arbitrary order on state
variables within each SCC. For every edge from one SCC to
another, i.e., where some state variable in the second SCC depends on some
state variable in the first, $s_1$ precedes $s_2$ in the order, 
where $s_2$ is the minimal element in the second SCC and $s_1$ 
is the maximal element in the first SCC. The state tests are then ordered
based on the order of state variables.

We translate a program to an xFDD using the $\fdd$ function (Figure~\ref{fig:fdd-syntax}), which
translates small parts of a program directly to xFDDs.
Composite programs get recursively translated and then
composed using a corresponding \cam{composition operator for \xFDDs}: we use 
$\oplus$ for $p + q$, 
$\odot$ for $p$ ; $q$, and $\ominus$ for $\neg p$. 
Figure~\ref{fig:fdd-normalization} gives a high-level 
definition of the semantics of these operators.
For example, $d_1 \oplus d_2$
tries to merge similar test 
nodes recursively by merging their true branches together and false ones together. 
If the two tests are not the same and $d_1$'s test comes first in the total order, both of its subtrees
are merged recursively with $d_2$. The other case is similar.
$d_1 \oplus d_2$ for leaf nodes is the union of their action sets.

The hardest case is surely for $\odot$, where we try to add in an
action sequence $as$ to an xFDD $(t ~?~ d_1 : d_2)$. 
Suppose we want to compose
$f \gets v_1$ with $(f = v_2~?~d_1 : d_2)$. The result of this xFDD composition
should behave as if we first do the update and then the condition on $f$. If 
$v_1 = v_2$, the composition should continue only on $d_1$, and if not,
only on $d_2$. Now let's look at a similar example including state, composing
$s[srcip] \gets e_1$ with $(s[dstip] = e_2 ~?~ d_1 : d_2)$. 
If $srcip$ and $dstip$ are equal (rare but not impossible) and $e_1$ and $e_2$
always evaluate to the same value, 
then the whole composition reduces to just $d_1$.
The field-field tests are introduced to let us answer these equality questions, and that is why
they always precede state tests in the tree. 
The trickiness in the algorithm comes from generating proper
field-field tests, by keeping
track of the information in the xFDD,
to \cam{properly answer the equality tests of interest}.
The full algorithm is given in appendix~\ref{app:st-dep}.

Note that the actual definition of the \xFDD composition operators is a bit more involved than 
the one in Figure~\ref{fig:fdd-normalization} as we have to make sure, while composing
FDDs, that the resulting FDD is \emph{well-formed}. An FDD is defined to be well-formed 
if its tests conform to the pre-defined total order ($\fddl$) and do not contradict the previous 
tests in the FDD. 
Figure~\ref{fig:fdd-normalization-detailed} contains a more detailed 
definition of $\oplus$ as an example. 
To detect possible contradictions, we accumulate both the equalities
and inequalities implied by previous tests in an argument called $context$ 
and pass it through recursive calls to $\oplus$. Before applying $\oplus$ to the input FDDs,
we first run each of the FDDs through a function called \textproc{refine}, which removes
both redundant and contradicting tests from top of the input FDD based on the input $context$ 
until it reaches a non-redundant and non-contradicting test. After both input FDDs are
``refined'', we continue with the merge as before.

Finally, recall from \cref{sec:language} that Inconsistent use of state variables 
is prohibited by the language semantics when composing programs.
We enforce
the semantics by 
looking for these violations while merging the xFDDs of composed
programs
and raising a compile error if the final xFDD contains 
a leaf with parallel updates to the same state variable. 

\subsection{Packet-State Mapping}
\label{sec:psm}

\newcommand{\psm}{\textproc{PSM}}

For a given program $p$, the corresponding xFDD $d$ offers
an explicit and complete specification of the way $p$
handles packets. We analyze $d$, using an algorithm called
\emph{packet-state mapping}, to determine which 
\emph{flows} use which states. This information is further used in the 
optimization problem (\cref{sec:milp}) to decide the correct 
routing for each flow.
Our default definition of a flow is those packets that travel between
any given pair of ingress/egress ports in the OBS, though
we can use other notions of flow (see
\cref{sec:milp}).
Traversing from $d$'s root down to the action sets at $d$'s leaves, we
can gather information associating each flow with the set of state
variables read or written. 
See appendix~\ref{app:fdd-seq} for the full algorithm.

Furthermore, the operators can give hints to the compiler by specifying
their network \emph{assumptions} in a separate policy:

{
	\mdfsetup{
		skipabove=2mm,
		skipbelow=3mm,
		rightmargin=0.3cm,
		leftmargin = 0.3cm,
	}
	\begin{mdframed}
		\raggedright
		\codeb[scriptsize]{
			\hspace{-2mm}assumption = (\match{srcip}{10.0.1.0/24} \& \match{inport}{1}) \\
			\hspace{1.45cm} + (\match{srcip}{10.0.2.0/24} \& \match{inport}{2}) \\
			\hspace{1.45cm} + ... \\
			\hspace{1.45cm} + (\match{srcip}{10.0.6.0/24} \& \match{inport}{6}) \\
		} 
	\end{mdframed}
}
 
We require the assumption policy to be a predicate over packet header 
fields, only passing the packets that match the operator's assumptions. 
\codeb{assumption} is then sequentially composed with the rest of the program,
enforcing the assumption by dropping packets that do not match the assumption.
Such assumptions benefit the packet-state mapping. Consider our 
example xFDD in Figure~\ref{fig:ex-fdd3}. 
Following the xFDD's tree structure, we can infer that all the packets
going to port 6 need all the three state variables in \Tunnel. We can 
also infer that all the packets coming from the 10.0.6.0/24 subnet need
\codeb{orphan} and \codeb{susp-client}. However, there is nothing in the program to
tell the compiler that these packets can only enter the network from port 6. 
Thus, the above assumption policy can help the compiler to identify this relation
and place state more efficiently.

\subsection{State Placement and Routing}
\label{sec:milp}

\cam{At this stage, the compiler has enough information to fill in the details abstracted away from the programmer:
where and how each state variable should be placed, and how the traffic should be routed in the network.
There are two general approaches for deciding state placement and routing. One is to keep
\emph{each} state variable at one location and route the traffic through the 
state variables it needs. The other is to keep multiple copies of the same state variable on different switches and 
partition and route the traffic through them.
The second approach requires mechanisms to keep different copies of the same state variable consistent. However, it is not possible to provide strong consistency guarantees when distributed updates are made on a packet-by-packet basis at line rate. Therefore, we chose the first approach, which locates each state variable at one physical switch.}

To decide state placement and routing, we
generate an optimization problem, a \emph{mixed-integer linear
  program} (MILP) that is an extension of the multi-commodity flow
linear program. The MILP has three key inputs: the concrete network
topology, the state dependency graph $G$, and the
packet-state mapping, and two key outputs: routing
and state placement (Table~\ref{tab:milp-inout}). 
\cam{Since route selection depends on state placement and each 
state variable is constrained to one physical location, we need to make sure
the MILP picks \emph{correct} paths without degrading network
performance. 
Thus, the MILP minimizes the sum of link utilization in the network as a measure of congestion. 
However, other objectives or constraints are conceivable 
to customize the MILP to other kinds of performance requirements.}

\snaptitle{Inputs.}
The topology is defined in terms of the following inputs to the MILP: 
\begin{inlinelist}
\item the nodes, some distinguished as edges (ports in OBS),
\item expected traffic $d_{uv}$ for every pair of edge nodes $u$ and $v$, and
\item link capacities $c_{ij}$ for every pair of nodes $i$ and $j$.
\end{inlinelist}
State dependencies in $G$ are translated
into input sets $dep$ and $tied$. $tied$ 
contains pairs of state variables which are in the same SCC
in $G$, and must be co-located.
$dep$ identifies state variables with dependencies that do not need to be co-located;
in particular, $(s,t) \in dep$ when $s$ precedes $t$ in variable ordering, and they are not
in the same SCC in $G$.
The packet-state mapping is used as
the input variables $S_{uv}$, identifying the set of
state variables needed on flows between nodes $u$ and $v$.

\begin{table}[t!]
\small
\centering
\begin{tabular}{|c|l|}\hline
\multicolumn{1}{|c}{\bf Variable} & \multicolumn{1}{|c|}{\bf Description}\\ \hline
$u, v$ & \text{edge nodes (ports in OBS)} \\
$n$ & \text{physical switches in the network}\\
$i, j$ & \text{all nodes in the network} \\
$d_{uv}$ & \text{traffic demand between $u$ and $v$} \\
$c_{ij}$ & \text{link capacity between $i$ and $j$} \\
$dep$ & \text{state dependencies} \\
$tied$ & \text{co-location dependencies }\\
$S_{uv}$ & \text{state variables needed for flow $uv$} \\ 
\hline
$R_{uvij}$ & fraction of $d_{uv}$ on link $(i,j)$\\
$P_{sn}$ & 1 if state $s$ is placed on $n$, 0 otherwise\\
$P_{suvij}$ & $d_{uv}$ fraction on link ($i,j$) that has passed $s$ \\
\hline
\end{tabular}
\caption{Inputs and outputs of the optimization problem. 
\label{tab:milp-inout}}
\end{table}

\snaptitle{Outputs and Constraints.}  The routing outputs are variables
$R_{uvij}$, indicating what fraction of the flow from edge node $u$
to $v$ should traverse the link between nodes $i$ and $j$.  The
constraints on $R_{uvij}$ (left side of
Table~\ref{tab:milp-constraints}) follow the multi-commodity flow
problem closely, with standard link capacity and flow conservation
constraints, and edge nodes distinguished as sources and sinks of
traffic.

State placement is determined by the variables $P_{sn}$, which
indicate whether the state variable $s$ should be placed on the physical switch
$n$.
Our constraints here are more unique to our setting.
First, every state variable $s$ can be placed on exactly one
switch, a choice we discussed earlier in this section.
Second, we must ensure that flows that need a given state variable $s$ traverse
that switch.
Third, we must ensure that each flow traverses states in the order
specified by the $dep$ relation; this is what the variables
$P_{suvij}$ are for. We require that $P_{suvij} = R_{uvij}$ when the
traffic from $u$ to $v$ that goes over the link $(i, j)$ has already
passed the switch with the state variable $s$, and zero otherwise.  If
$dep$ requires that $s$ should come before some other state variable $t$---and
if the $(u, v)$ flow needs both $s$ and $t$---we can use $P_{suvij}$
to make sure that the $(u, v)$ flow traverses the switch with $t$ only
after it has traversed the switch with $s$ (the last state constraint in
Table~\ref{tab:milp-constraints}).
Finally, we must make sure that state variables $(s,t) \in tied$ are located on
the same switch.
\cam{Note that only state variables that are \emph{inter-dependent} are required
to be located on the same switch. Two variables $s$ and $t$ are inter-dependent if 
a read from $s$ is required before a write to $t$ \emph{and vice versa}.
Placing them on different switches will result in a forwarding 
loop between the two switches which is not desirable in most networks.
Therefore, in order to synchronize reads and writes to inter-dependent variables correctly, 
they are always placed on the same switch.}

\cam{Although the current prototype chooses the same path for the traffic between the same
ports,} the MILP can be configured to decide paths for more fine-grained notions of flows.
Suppose packet-state mapping finds that only packets with $srcip=x$
need state variable $s$. We refine the MILP input to have two edge nodes per
port, one for traffic with $srcip=x$ and one for the rest, so
the MILP can choose different paths for them.

\cam{Finally, the MILP makes a \emph{joint} decision for state placement and routing. 
Therefore, path selection is tied to state placement. To have more freedom in picking
forwarding paths, one option is to first use common traffic engineering techniques to decide
routing, and then optimize the placement of state variables with respect to the selected paths.
However, this approach may require replicating state variables and maintaining consistency 
across multiple copies, which as mentioned earlier, is not possible at line rate for distributed 
packet-by-packet updates to state variables.
}

\subsection{Generating Data-Plane Rules}
\label{sec:rulegen}

Rule generation happens in two phases and combines
information from the xFDD and MILP to configure the
network switches.
We assume each packet is
augmented with a \lang-header upon entering the network, 
which contains its original OBS inport and future outport, 
and the id of the last processed xFDD node, the purpose of 
which will be explained shortly.
This header is stripped off \cam{by the egress switch} when the packet exits the network.
We use \codeb{\Tunnel[footnotesize];assign-egress} from \cref{sec:example} as a running example,
with its xFDD in Figure~\ref{fig:ex-fdd3}. For the sake of the example, 
we assume that all the state variables are stored on $C_6$
instead of $D_4$.

\begin{table}[t!]
\scriptsize
\centering
\begin{tabular}{|>{$}l<{$}|>{$}l<{$}|}
\hline
\text{\textbf{Routing Constraints}} & \text{\textbf{State Constraints}} \\
\hline
& \sum_{n} P_{sn} = 1\\
\sum_j R_{uvuj} = 1 & \forall u, v.~ \forall s \in S_{uv}.~\sum_i R_{uvin} \ge P_{sn}\\
\sum_i R_{uviv} = 1& \forall (s, t)\in tied.~P_{sn} = P_{tn}\\
\sum_{u,v} R_{uvij}d_{uv} \le c_{ij} & P_{suvij} \leq R_{uvij} \\
\sum_i R_{uvin} = \sum_j R_{uvnj} & P_{sn} + \Sigma_{i} P_{suvin} = \Sigma_{j} P_{suvnj} \\
\sum_i R_{uvin} \leq 1 & \forall s \in S_{uv}.~P_{sv} + \sum_i P_{suviv} = 1 \\
& P_{sn} + \Sigma_i P_{suvin} \ge P_{tn} \\
\hline
\end{tabular}
\caption{Constraints of the optimization problem.}
\label{tab:milp-constraints}
\end{table}

In the first phase, we break the xFDD down into `per-switch' xFDDs, since not every switch
needs the entire xFDD to process packets.
Splitting the xFDD is straightforward given placement
information: stateless tests and actions can happen anywhere, but
reads and writes of state variables must happen on switches storing them.
For example, edge switches ($I_1$ and $I_2$, and $D_1$ to $D_4$)
only need to process packets up to the state tests, e.g., tests 3 and
8, and write the test number in the packet's \lang-header showing
how far into the xFDD they progressed.
Then, they send the packets to $C_6$, which has the corresponding state
variables, \codeb{orphan} and \codeb{susp-client}. $C_6$, on the other
hand, does not need the top part of the xFDD. It just needs the
subtrees containing its state variables to continue processing the
packets sent from the edges.
The per-switch xFDDs are then translated to switch-level configurations, by
a straightforward traversal of the xFDD (See \cref{sec:implementation}).

In the second phase, we generate a set of match-action 
rules that take packets through the paths decided by the MILP. These
paths comply with the state ordering used in the xFDD, 
thus they get packets to switches with the right states in the right order.
Note that packets contain the path identifier (the OBS inport and outport, 
$(u, v)$ pair in this case) 
and the ``routing'' match-action rules are generated in terms of this identifier
to forward them on the correct path.
Additionally, note that it may not always be possible to decide the egress
port $v$ for a packet upon entry if its outport depends on state. 
We observe that in that case, all the paths for possible
outports of the packet pass the state variables it needs. We 
\cam{load-balance over} these paths in proportion to their capacity and show, in 
appendix~\ref{appendix}, that traffic on
these paths remains in their capacity limit.

\cam{To see an example of how packets are handled by generated rules,}
consider a DNS response with source IP 10.0.1.1 and destination IP
10.0.6.6, entering the network from port 1. The rules on $I_1$
process the packet up to test 8 in the xFDD, tag the packet with the
path identifier (1, 6) and number 8. The packet is then sent to $C_6$.
There, $C_6$ will process the packet from test 8, update state variables 
accordingly, and send the packet to $D_4$ to exit the network from port 6.

\section{Implementation}\label{sec:implementation}

The compiler is mostly implemented in Python, except for 
the state placement and routing phase (\cref{sec:milp}) 
which uses the Gurobi Optimizer~\cite{gurobi} to solve the MILP.
The compiler's output for each switch 
is a set of switch-level instructions
in a low-level language called \nohyphens{NetASM}~\cite{netasm}, which
comes with a software switch capable of executing 
those instructions.
NetASM is an assembly language for programmable data planes 
designed to serve as the ``narrow waist''  between high-level 
languages such as \lang, and NetCore\cite{NetCore}, 
and programmable switching architectures such as RMT~\cite{RMT}, FPGAs, 
network processors and Open vSwitch.

As described in \cref{sec:rulegen}, each switch processes the packet
by its customized per-switch \xFDD, and then forwards it based on
the fields of the \lang-header using a match-action table.
To translate the switch's \xFDD to \nohyphens{NetASM} instructions, we 
traverse the \xFDD and generate a \emph{branch} instruction for each
test node, which jumps to the instruction of either the true or false branch
based on the test's result.
Moreover, we generate instructions to create two tables for each
state variable, one for the indices and one for the values.
In the case of a state test in the \xFDD, we first retrieve the value corresponding
to the index that matches the packet, and then perform the branch.
For \xFDD leaf nodes, we generate \emph{store} instructions that modify 
the packet fields and state tables accordingly. 
Finally, we use NetASM support for atomic execution of multiple instructions
to guarantee that operations on state tables happen atomically.

While NetASM was useful for testing our compiler, any programmable device
that supports match-action tables, 
branch instructions, and stateful operations 
can be a \lang target.
The prioritized rules in match-action tables, for instance, are effectively branch instructions. Thus,
one can use multiple match-action tables to implement \xFDD in the data plane, generating 
a separate rule for each path in the \xFDD.
Several emerging switch interfaces support stateful 
operations~\cite{p4, openstate, pof, openvswitch}. We discuss possible 
software and hardware implementations for \lang stateful
operations in~\cref{sec:discussion}.
\section{Evaluation}
\label{sec:evaluation}

\newcommand{\DEP}{P1}
\newcommand{\XFDD}{P2}
\newcommand{\PSM}{P3}
\newcommand{\MILPCR}{P4}
\newcommand{\MILPSOL}{P5}
\newcommand{\RULGEN}{P6}

This section evaluates \Lang in terms of language expressiveness and 
compiler performance.

\subsection{Language Expressiveness}

\begin{table}[t!]
	\centering
	\scriptsize
	\begin{tabular}{| l | l |} \cline{2-2}
\multicolumn{1}{c|}{} & 
  \multicolumn{1}{c|}{\bf Application} \\ \hline
\multirow{5}{1.3cm}{Chimera~\cite{chimera}}& \# domains sharing the same IP address \\ 
& \# distinct IP addresses under the same domain \\ 
& DNS TTL change tracking \\ 
& DNS tunnel detection\\ 
& Sidejack detection\\ 
& Phishing/spam detection\\ \hline
\multirow{6}{1cm}{FAST~\cite{FAST}}& Stateful firewall\\ 
& FTP monitoring\\ 
& Heavy-hitter detection\\ 
& Super-spreader detection\\ 
& Sampling based on flow size\\ 
& Selective packet dropping (MPEG frames)\\ 
& Connection affinity\\ \hline
\multirow{4}{1cm}{Bohatei~\cite{bohatei}}& SYN flood detection \\ 
& DNS amplification mitigation\\ 
& UDP flood mitigation\\ 
& Elephant flows detection\\ \hline
\multirow{2}{1cm}{Others} & Bump-on-the-wire TCP state machine \\ 
& Snort flowbits~\cite{Snort}\\ \hline
\end{tabular}
\caption{Applications written in \Lang.}
\label{fig:example-list}
\end{table}

We 
have implemented several stateful network functions (Table~\ref{fig:example-list}) that are typically relegated to middleboxes in \lang. Examples were taken from the
Chimera~\cite{chimera}, FAST~\cite{FAST}, and Bohatei~\cite{bohatei} systems.
The code can be found in appendix~\ref{app:examples}.
Most examples use protocol-related fields in fixed
packet-offset locations, which are parsable by emerging programmable parsers.
Some fields 
require session reassembly. 
However, this is orthogonal to the
language expressiveness; as long as these fields are available to the
switch, they can be used in \Lang
programs.
To make them available, 
one could extract these fields by placing a ``preprocessor''
before the switch pipeline, similar to middleboxes. 
For instance, Snort~\cite{Snort} uses preprocessors 
\cam{to extract fields for use in the detection engine.}

\subsection{Compiler Performance}

The compiler goes through several phases upon the system's cold start, 
yet most events require only some of them. 
Table~\ref{tab:phase-per-change} summarizes these
phases and their sensitivity to network and policy changes. 

\snaptitle{Cold Start.} When the very first
program is compiled, the compiler
goes through all phases, including MILP model 
creation, which happens \emph{only once} in the lifetime
of the network.
Once created, the model supports incremental additions and modifications of 
variables and constraints in a few milliseconds. 

\snaptitle{Policy Changes.} Compiling a \emph{new} program requires executing the three program analysis phases and rule generation as well as 
\emph{both} state placement and routing, which are
decided using the MILP in~\cref{sec:milp}, denoted by ``ST''. 
Policy changes become considerably
\emph{less frequent}~(\cref{subsec:compiler-overview}) 
since most dynamic changes
are captured by the state variables that reside on the data plane.
The policy, and consequently switch configurations,
\emph{do not} change upon state changes.
Thus, we expect policy changes to happen infrequently, and be planned in
advance. The Snort rule set, for instance, gets updated every few
days~\cite{snort-blog}.

\snaptitle{Topology/TM Changes.}
Once the policy is compiled, we fix the decided
state placement, and only re-optimize routing in response to network 
events such as failures. 
For that, we formulated a variant of ST, denoted as ``TE'' (traffic engineering), that
receives state placement as input, and decides forwarding paths while 
satisfying state requirement constraints. 
\cam{\emph{We expect TE to run every few minutes}} since in a typical network,
the traffic matrix is fairly stable and traffic engineering happens on the timescale of \emph{minutes}~\cite{google-b4, tm-reloaded, nucci2005problem,
  suchara2011network}. 

\begin{table}[t!]
\setlength{\belowcaptionskip}{-2mm}
\setlength\tabcolsep{3pt}
\scriptsize
\centering
\begin{tabular}{| l | l | l | c | c | c |}
\hline
\textbf{ID} & \multicolumn{2}{c|}{\textbf{Phase}} & \begin{tabular}{@{}c@{}}\textbf{\tiny Topo/TM} \\ \textbf{\tiny Change}\end{tabular} & 
\begin{tabular}{@{}c@{}}\textbf{\tiny Policy} \\ \textbf{\tiny Change}\end{tabular} & \begin{tabular}{@{}c@{}} \textbf{\tiny Cold} \\ \textbf{\tiny Start}\end{tabular} \\
\hline
\DEP & \multicolumn{2}{c |}{State dependency} & - & \checkmark & \checkmark\\
\XFDD & \multicolumn{2}{c |}{xFDD generation} & - & \checkmark & \checkmark\\
\PSM & \multicolumn{2}{c |}{Packet-state map} & - & \checkmark & \checkmark \\
 \hline
\MILPCR & \multicolumn{2}{c |}{MILP creation} & - & - & \checkmark\\ \hline
\multirow{2}{*}{\MILPSOL} & \multirow{2}{*}{\begin{tabular}{@{}c@{}} \text{MILP} \\ \text{solving}\end{tabular}} & 
					   \begin{tabular}{@{}c@{}} \text{State placement } \\ \text{and routing (ST)}\end{tabular} & - & \checkmark & \checkmark\\ \cline{3-6}
& & Routing (TE) & \checkmark &  - & -\\
\hline
\RULGEN & \multicolumn{2}{c |}{Rule generation} & \checkmark & \checkmark & \checkmark\\
\hline
\end{tabular}
\caption{Compiler phases. For each scenario, phases that get executed are checkmarked.}
\label{tab:phase-per-change}
\end{table}

\subsubsection{Experiments}
We evaluated performance
based on applications listed in Table~\ref{fig:example-list}.
Traffic matrices are synthesized
using a gravity model~\cite{tm-synthesis}.  We used an
Intel Xeon E3, 3.4 GHz, 32GB server, and PyPy 
compiler \cite{pypy}. 

\begin{table}
\scriptsize
\centering
\begin{tabular}{|l|c|c|c|} \hline
\textbf{Topology} & \textbf{\# Switches} & \textbf{\# Edges} & \textbf{\# Demands} \\ \hline
Stanford & 26 & 92 & 20736 \\
Berkeley & 25 & 96 & 34225 \\
Purdue & 98 & 232 & 24336 \\
\hline
AS 1755 & 87 & 322 & 3600 \\
AS 1221 & 104 & 302 & 5184 \\
AS 6461 & 138 & 744 & 9216 \\
AS 3257 & 161 & 656 & 12544 \\
\hline
\end{tabular}
\caption{Statistics of evaluated enterprise/ISP topologies.}
\label{tab:enterprise-summary}
\end{table}

\begin{figure*}[t!]
\setlength{\abovecaptionskip}{4mm}
\setlength{\belowcaptionskip}{-2mm}
\begin{minipage}[t]{0.32\textwidth}
\captionsetup{justification=centering, font=scriptsize}
\includegraphics[width= .9\linewidth]{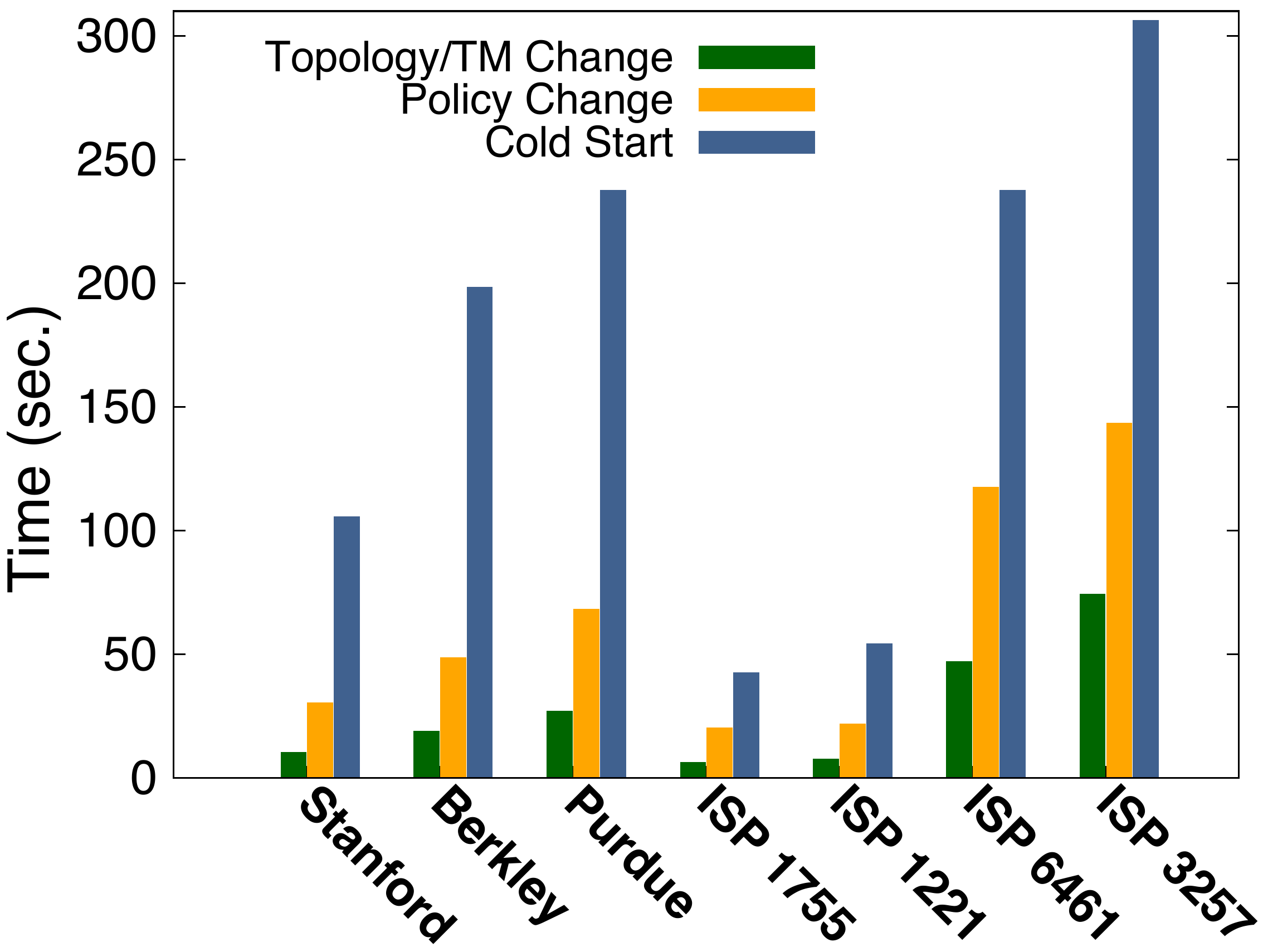}
\caption{Compilation time of \Tunnel[scriptsize] with routing on enterprise/ISP networks.\label{fig:enterprise-te}}
\end{minipage} \hfill
\begin{minipage}[t]{0.32\textwidth}
\captionsetup{justification=centering, font=scriptsize}
  \includegraphics[width=0.9\linewidth]{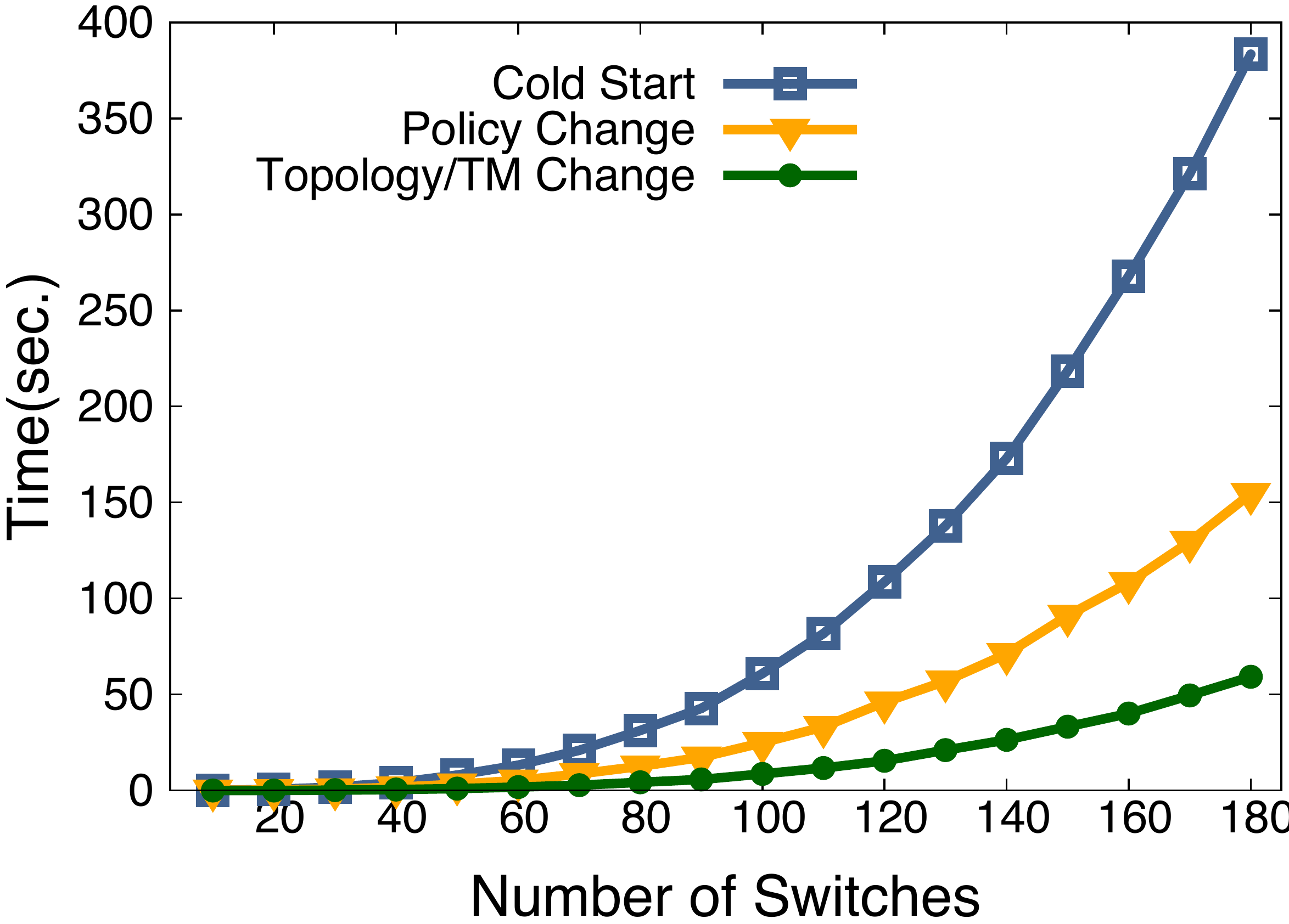}
  \caption{Compilation time of \Tunnel[scriptsize] with routing on IGen topologies.}
  \label{fig:eval_topo}
\end{minipage} \hfill
\begin{minipage}[t]{0.32\textwidth}
\captionsetup{justification=centering, font=scriptsize}
  \includegraphics[width= .9\linewidth]{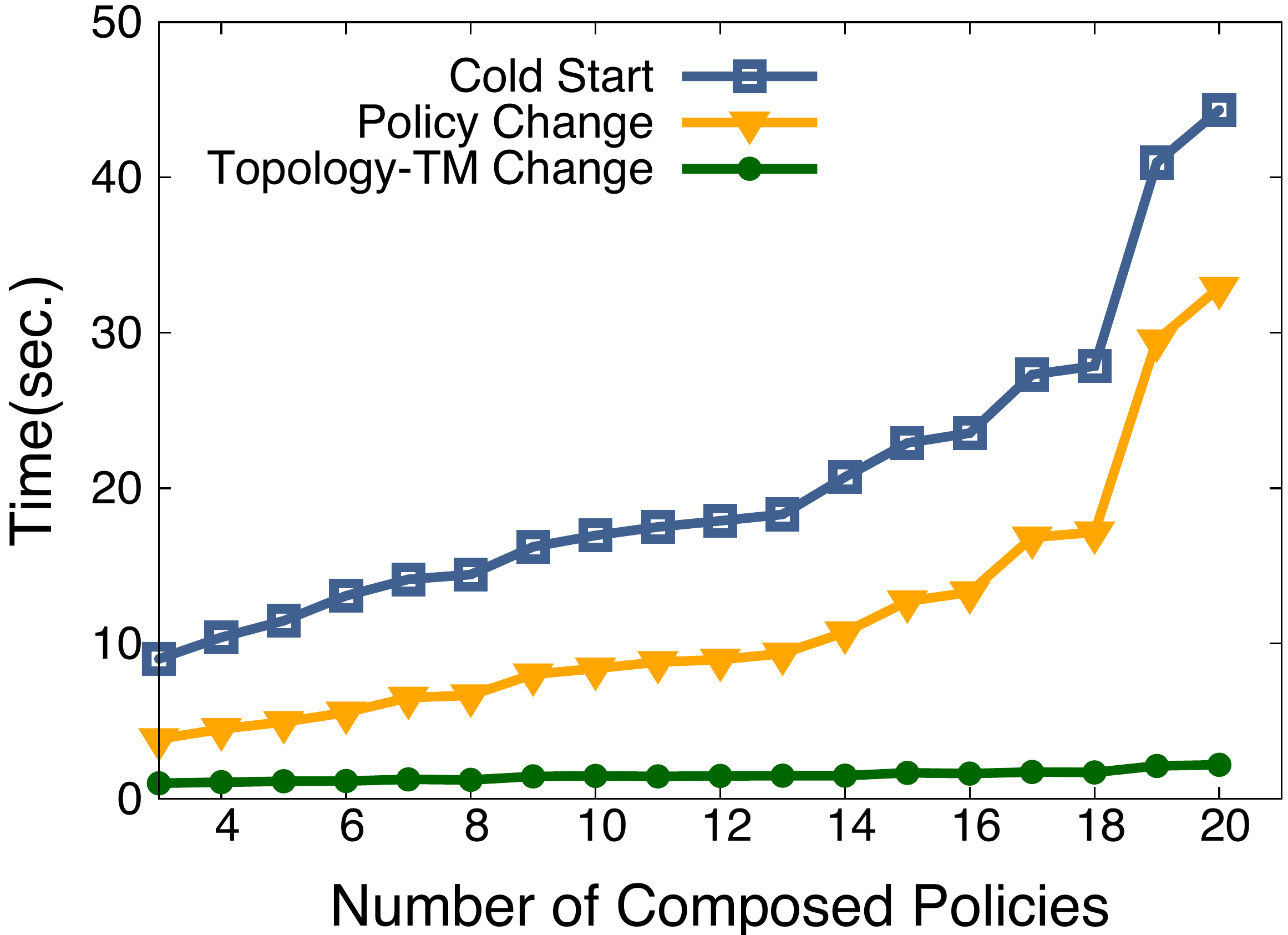}
  \caption{Compilation time for policies from Table~\ref{fig:example-list} incrementally
  		composed on a 50-switch network.}
 \label{fig:eval_state}
\end{minipage}
\caption{Compiler runtimes for scenarios in Table~\ref{tab:phase-per-change} on various policies and topologies. Once compiled for the first time (cold start, policy change), 
a policy reacts to traffic using its state variables. Topology/TM changes result in reoptimizing forwarding paths.}
\end{figure*}

\begin{table}
\centering
\scriptsize
\begin{tabular}{| l | c | c | c | c | c |} \cline{2-6}
\multicolumn{1}{c|}{} &  \multirow{2}{*}{\textbf{\DEP-\XFDD-\PSM} (s)} & 
				     \multicolumn{2}{c|}{\textbf{\MILPSOL} (s)}&
				     \multirow{2}{*}{\textbf{\RULGEN}(s)} & 
				     \multirow{2}{*}{\textbf{\MILPCR} (s)} \\

\cline{3-4}
\multicolumn{1}{c|}{} & & \textbf{ST} & \textbf{TE} & & \\
\hline 
Stanford	& 1.1	 & 29 & 10 & 0.1 & 75 \\
Berkeley     & 1.5 & 47 & 18 & 0.1 & 150\\
Purdue	& 1.2 & 67 & 27 & 0.1 & 169\\ \hline 
AS 1755	& 0.6	 & 19 & 6 & 0.04 & 22 \\ 
AS 1221	& 0.7	 & 21 & 7 & 0.04 & 32 \\  
AS 6461	& 0.8	 & 116 & 47 & 0.1 & 120 \\ 
AS 3257	& 0.9	 & 142 & 74 & 0.2 & 163 \\ \hline
\end{tabular}
\caption{Runtime of compiler phases when compiling \Tunnel[scriptsize] with routing on enterprise/ISP topologies.}
\label{tab:enterprise-results}
\end{table}

\snaptitle{Topologies.} 
We used a set of three campus
networks and four inferred ISP topologies from
RocketFuel~\cite{rocketfuel} (Table~\ref{tab:enterprise-summary}).\footnote{\footnotefont The publicly available
  Mininet instance of Stanford campus topology has
  10 extra dummy switches to implement multiple links between two routers.} For ISP networks, we considered 70\% of the
switches with the lowest degrees as edge switches to form OBS external
ports.  
The ``\# Demands'' column shows the number
of distinct OBS ingress/egress pairs. We assume directed
links.
Table~\ref{tab:enterprise-results} shows  compilation time for the DNS tunneling example (\cref{sec:example}) on each network, broken down by compiler phase. 
Figure~\ref{fig:enterprise-te} compares the compiler runtime for different
scenarios, combining the runtimes of phases relevant for each. 

\snaptitle{Scaling with topology size.} 
We synthesize networks
with 10--180 switches using IGen~\cite{igen}. In each network, 70\% of 
the switches with the lowest degrees are chosen as edges and the DNS tunnel policy is
compiled with that network as a target. Figure~\ref{fig:eval_topo} shows
the compilation time for different scenarios, combining the  
 runtimes of phases relevant for each. 
Note that by increasing the topology size,
the policy size also increases in the \codeb{assign-egress} and \codeb{assumption} parts.

\snaptitle{Scaling with number of policies.} 
\cam{The performance of several phases of the compiler, specially \xFDD generation, is a function of
the size and complexity of the input policy. Therefore, we 
evaluated how the compiler's performance scales with policy size using the example 
programs from Table~\ref{fig:example-list}. Given that these programs are taken from recent
papers and tools in the literature~\cite{chimera, FAST, bohatei, Snort}, we believe they form
a fair benchmark for our evaluation. Except for TCP state machine, the example programs are similar
in size and complexity to the DNS tunnel example (\cref{sec:example}).
We use the 50-switch network from the previous experiment and start with the first program in
Table~\ref{fig:example-list}. 
We then gradually increase the size of the final policy by combining this program with more programs from Table~\ref{fig:example-list} using the parallel composition operator.  
Each additional component program affects traffic destined to a separate egress port.

Figure~\ref{fig:eval_state} depicts the compilation time as a function of the number of 
components from Table~\ref{fig:example-list} that form the final policy.}
The $10$-second jump from 18 to 19 takes place when the TCP state machine
policy is added, which is considerably more complex than others.
\cam{The increase in the compilation time mostly comes from the \xFDD generation phase.
In this phase, the composed programs are transformed
into separate \xFDDs, which are then combined to form the \xFDD for the whole policy (\cref{sec:fdds}). 
The cost of xFDD composition depends on the size of the operands, so as more components are put together, the cost grows. 
The cost may also depend on the order of \xFDD composition.
Our current prototype composes \xFDDs in the same order as the programs themselves are composed
and leaves finding the optimal order to compose \xFDDs to future work. 
}

\cam{The last data point in Figure~\ref{fig:eval_state} shows the compilation time of a policy composed of
all the 20 examples in Table~\ref{fig:example-list}, with a total of 35 state variables. These policies are
composed using parallel composition, which does not introduce read/write dependencies between
state variables. Thus, the dependency graph for the final policy is a collection of the
dependency graphs of the composed policies. 
Each of the composed policies affects the traffic to a separate egress
port, which is detected by the compiler in the packet-state mapping phase. 
Thus, when compiled to the 50-switch network, state variables for each policy are placed on the switch
closest to the egress port whose traffic the policy affects. If a policy were to affect a larger
portion of traffic, e.g., the traffic of a set of ingress/egress ports, \lang would place state variables
in an optimal location where the aggregated traffic of interest is passing through. }
 
\subsubsection{Analysis of Experimental Results}

\emph{Creating the MILP takes longer than solving it}, in most cases,
and much longer than other phases. Fortunately, this
is a \emph{one-time} cost. After creating the MILP instance,
incrementally adding or removing variables and constraints (as the
topology and/or state requirements change) takes just a few milliseconds.

\emph{Solving the ST MILP unsurprisingly takes longer as compared to the
  rest of the phases} when topology grows. It takes $\scriptsize \sim$ 2.5
minutes for the biggest synthesized topology and $\scriptsize \sim$ 2.3 minutes for
the biggest RocketFuel topology.  The curve is close to
exponential as the problem is inherently computationally
hard. However, this phase takes place only  
in cold start or upon a \emph{policy} change, which  are infrequent 
and planned in advance. 

\emph{Re-optimizing routing
with fixed state placement is much faster}.  In response to network events
(e.g., link failures), TE MILP can
recompute paths in around a minute across all our experiments,
\cam{\emph{which is the timescale we initially expected for this phase} as it runs
in the topology/TM change scenarios}. 
Moreover, it can be used even on \emph{policy} changes,
if the user settles for a sub-optimal state placement using
heuristics rather than ST MILP. We plan to explore such 
heuristics.  

Given the kinds of events that require complete (policy change) or
partial (network events) recompilation, we believe that our compilation techniques 
meet the requirements of enterprise networks and medium-size ISPs.
Moreover, if needed,
our compilation procedure could be combined with 
traffic-engineering techniques once the state placement is decided, to
avoid re-solving the original or even TE MILP on small timescales.

\section{Discussion}
\label{sec:discussion}

This section discusses data-plane implementation strategies
for \lang's stateful operations, how \lang relates to middleboxes,
and possible extensions to our techniques to enable a broader range 
of applications.

\subsection{Stateful Operations in the Data Plane}
A state variable (array) in \Lang is a key-value mapping, or a \emph{dictionary}, 
on header fields, persistent across multiple packets. 
When the key (index) range is small, it is feasible to pre-allocate all the memory
the dictionary needs and implement it using an array.
A large but \textit{sparse} dictionary 
can be implemented using a \textit{reactively}-populated table, 
similar to a MAC learner table.
It contains a single default entry in the beginning,
and as packets fly by and change the state variable, 
it \textit{reactively} adds/updates the corresponding entries.

In software, there are efficient techniques to implement a dictionary in either approach, and some software switches already support similar reactive ``learning'' operations, either atomically~\cite{netasm} or
with small periods of inconsistency~\cite{openvswitch}. 
The options for current hardware are: 
\begin{inlinelist}
\item arrays of registers, which are already supported in 
emerging switch interfaces~\cite{p4}. 
They can be used to implement small dictionaries, as well as
Bloom Filters and hash tables as sparse dictionaries.
In the latter case, it is possible for two different keys to hash to the same dictionary entry. 
However, there are applications such as load balancing and flow-size-based sampling
that can tolerate such collisions \cite{FAST}.
\item Content Addressable Memories (CAMs) are typically present in today's 
hardware switches and can be modified by a software agent running on the switch.
Since CAM updates triggered by a packet are not immediately available to
the following packets,  
it may be used for applications that tolerate small periods of state inconsistency, such as 
a MAC learner, DNS tunnel detection, and others from Table~\ref{fig:example-list}. 
\end{inlinelist}
Our NetASM implementation  (\cref{sec:implementation}) takes the CAM-based approach.
NetASM's software switch supports atomic updates to the tables in the data plane and therefore can
perform \emph{consistent} stateful operations. 

\cam{At the time of writing this paper, we are not aware of any hardware switch
that can implement an \emph{arbitrary} number of \lang's stateful operations both at \emph{line rate}
and with \emph{strong consistency}. Therefore, we use \nohyphens{NetASM's} low-level primitives as the 
compiler's backend so that we can specify data-plane primitives that are required for an efficient and 
consistent implementation of \lang's operations. If one is willing to relax one of the above constraints
for a specific application, i.e., operating at line rate or strong consistency, 
it would be possible to implement \lang on today's switches.
If strong consistency is relaxed, CAMs/TCAMs can be programmed using languages such as P4~\cite{p4}
to implement \lang's stateful operations as described above.  
If line-rate processing is relaxed, one can use software switches, or programmable hardware switching 
devices such as ones in the OpenNFP project that allow insertion of Micro-C code extensions to P4 programs 
at the expense of processing speed~\cite{opennfp} or FPGAs.}

\subsection{\lang and Middleboxes}

Networks traditionally rely on middleboxes for 
advanced packet processing, including stateful functionalities.
However, advances in switch technology enable stateful packet processing
in the data plane, which naturally makes the switches capable of 
subsuming a subset of middlebox functionality.
\lang provides a \emph{high-level programming 
framework} to exploit this ability, hence, it is able to express a wide range of stateful programs 
that are typically relegated to middleboxes (see Table~\ref{fig:example-list} for examples). 
This helps the programmer to 
think about a single, explicit network policy, as opposed to a disaggregated, implicit network policy
using middleboxes, and therefore, get more control and customization over a variety of
simpler stateful functionalities.

This also makes \lang subject to similar challenges  
as managing stateful middleboxes.
For example, many network functions must observe all traffic pertaining to
a connection \emph{in both directions}.  
In SNAP, 
if traffic in both directions
uses a shared state variable, the MILP optimizer
forces traffic in both directions through the same node.
Moreover, previous work such as
Split/Merge~\cite{SplitMerge} and OpenNF~\cite{OpenNF} 
show how to migrate \emph{internal} state from one network
function to another, and Gember-Jacobson et
al.~\cite{GJ} manage to  migrate state without buffering 
packets at the controller. 
\lang currently focuses on static state placement. However,
since \lang's state variables are explicitly declared as part of the policy, rather
than hidden inside blackbox software,
\lang is well situated to adopt these algorithms to support
smooth transitions of state variables in dynamic state 
placement.
Additionally, the \lang compiler can easily analyze a program to
determine whether a switch modifies packet fields
to ensure correct traffic steering---something that is challenging today with 
blackbox middleboxes~\cite{FlowTags,Simple}.

While \lang goes a step beyond previous high-level languages
to incorporate stateful programming into SDN,
we neither claim that it is as expressive 
as all stateful middleboxes, nor that it can replace 
them. 
To interact with middleboxes, 
SNAP may adopt techniques such as 
FlowTags~\cite{FlowTags} or SIMPLE~\cite{Simple} to
direct traffic through middleboxs chains by
tagging packets to mark their progress.
\cam{Since \lang has its own tagging and steering to keep track
of the progress of packets through the policy's \xFDD, this adoption
may require integrating tags in the middlebox framework with \lang's tags. 
As an example, we will describe below how \lang and FlowTags can be used
together on the same network.}

\cam{In FlowTags, 
users specify which class of traffic should pass which chain of middleboxes under what conditions.
For instance, they can ask for web traffic to go to an intrusion detection system (IDS) after a firewall if
the firewall marks the traffic as suspicious. The controller keeps a mapping between the tags and the flow's original 
five tuple plus the contextual information of the last middlebox, e.g., suspicious vs. benign in the case of a firewall. The tags are used
for steering the traffic through the right chain of middleboxes and preserving the original information of the flow in case it is 
changed by middleboxes.}
\cam{To use FlowTags with \lang, we can treat middlebox contexts as state variables and
transform FlowTags policies to \lang programs. Thus, they can
be easily composed with other \lang policies. 
Next, we can fix the placement of middlebox state variables to the actual location of the middlebox 
in the network in \lang's MILP. This way, \lang's compiler can decide state placement and routing for 
\lang's own policies while making sure that the paths between different middleboxes in
the FlowTags policies exist in the network. Thus, steering happens using \lang-generated tags.
Middleboxes can still use tags from FlowTags to learn about flow's original information or the context of the previous middlebox.
}

Finally, we focus on \emph{programming} networks
but if
verification is of interest in future work, one might adopt techniques such as 
RONO~\cite{RONO} to verify isolation properties 
in the presence of stateful middleboxes.  In summary,
interacting with existing middleboxes is no harder or easier
in \lang than it is in other \emph{global} SDN languages, \cam{stateless or stateful}, such as 
NetKAT~\cite{netkat} or Stateful NetKAT~\cite{StatefulNetKAT}. 
\cam{ 
\subsection{Extending \lang}
\label{subsec:extensions}
\snaptitle{Sharding state variables.}
The MILP assigns each state variable to
\emph{one} physical switch to avoid the overhead
of synchronizing multiple instances of the same variable.
Still, distributing a state variable remains a valid option. For instance, the compiler
can partition $s[inport]$ into $k$ \emph{disjoint} state variables, each storing
$s$ for one port. The MILP can decide 
placement and routing as before, this time with the option of distributing partitions of $s$
with no concerns for synchronization. See appendix~\ref{app:sharding} for more details.

\snaptitle{Fault-Tolerance.}
\lang's current prototype does not implement any particular fault tolerance mechanism in case a switch holding a state variable fails.
Therefore, the state on the failed switch will be lost. 
However, this problem is not inherent or unique to \lang and will happen in existing solutions with middleboxes too if the state of the middlebox is not replicated. 
Applying common fault tolerance techniques to switches with state to avoid state loss in case of failure can be an interesting direction for future work.

\snaptitle{Modifying fields with state variables.}
An interesting extension to \lang is allowing a packet field
to be directly modified with the value of a state variable at a 
specific index:
\codeb{\modify{f}{s[e]}}.
This action can be used in applications such as NATs and proxies,
which can store connection mappings in state variables and modify 
packets accordingly as they fly by.
Moreover, this action would enable \lang programs to modify a field
by the output of an arbitrary function on a set of packet fields,
such as a hash function.
Such a function is nothing but a fixed mapping between input header fields and 
output values. Thus, when analyzing the program, the compiler can treat 
these functions as fixed state variables with the function's input fields as index
for the state variable
and place them on switches with proper capabilities when distributing
the program across the network.
However, adding this action results in 
complicated dependencies between program statements, which is 
interesting to explore as future work. 

\snaptitle{Deep packet inspection (DPI).} Several applications such as intrusion detection require 
searching the packet's payload for specific patterns. 
\Lang can be extended 
with an extra field called \emph{content}, containing the packet's payload.
Moreover, the semantics
of tests on the content field can be extended to match on regular expressions.
The compiler can also be modified to assign content tests 
to switches with DPI capabilities. 

 \snaptitle{Resource constraints.}
 \Lang's compiler
 optimizes state placement and routing for link utilization.
 However, other resources such as switch memory and processing power in terms of maximum number of 
 complicated operations on packets (such as stateful updates, increments, or decrements) may limit the possible computations on
 a switch.  
 An interesting direction for future work would be to augment the \lang compiler with the ability to optimize for these additional resources.

\snaptitle{Cross-packet fields.}
Layer 4-7 fields are useful for classifying flows in
stateful applications, but are often scattered across multiple physical packets.
Middleboxes typically perform session reconstruction to extract these fields. 
Although \Lang language is agnostic to the chosen set of fields, 
the compiler currently supports fields stored \emph{in the packet itself}
and the state associated with them. However, it may be interesting to explore
abstractions for expressing how multiple packets (e.g., in a session) can form
``one big packet'' and use its fields. 
The compiler can further
place sub-programs that use cross-packet fields on devices that are capable of reconstructing
the ``one big packet''.

\snaptitle{Queue-based policies.}
\Lang currently has no notion of queues and therefore, cannot be used to express queue-based
performance-oriented policies such as active queue management, queue-based load balancing, and 
packet scheduling.
There is ongoing research on finding the right set of primitives for expressing such
policies~\cite{pifo}, which is largely orthogonal and complementary to \lang's current goals.
}
\section{Related Work}\label{sec:related}

\snaptitle{Stateful languages.} 
Stateful NetKAT~\cite{StatefulNetKAT}, developed concurrently with \lang, 
is a stateful language
for ``event-driven'' network programming, which guarantees
consistent update when transitioning between configurations in response to events.
\lang source language is richer and exponentially more 
compact than stateful NetKAT as it
contains \emph{multiple arrays} (as opposed to one)
that can be indexed and updated by contents of \emph{packet headers} (as opposed to 
constant integers only).  
Moreover, they place multiple copies of state at the edge, proactively generate rules for all configurations, and optimize
for rule space,
while we distribute state
and optimize for congestion.
Kinetic~\cite{Kinetic} provides a per-flow state machine abstraction, 
and NetEgg~\cite{NetEgg} synthesizes stateful programs from user's examples.
However, they both keep the state at the controller.

\snaptitle{Compositional languages.}
NetCore \cite{NetCore}, and other similar languages \cite{pyretic, frenetic, netkat},
have primitives for tests and modifications on packet fields as well as composition operators 
to combine programs. \lang builds on these languages by adding primitives
for stateful programming (\cref{sec:language}).
To capture the joint intent of two policies, sometimes the programmer
needs to decompose them into their constituent pieces, and then reassemble them 
using \codeb{;} and \codeb{+}.
PGA \cite{pga} allows programmers to specify access control and service chain policies 
using graphs as the basic building block, and 
tackles this challenge by defining a new type of composition.
However, PGA does not have linguistic primitives for stateful programming, 
such as those that read and write the contents of global arrays.
Thus, we view \lang and PGA as complementary research
projects, with each treating different aspects of the language design
space. 

\snaptitle{Stateful switch-level mechanisms.}
FAST~\cite{FAST} and OpenState~\cite{openstate} 
propose flow-level state machines as a primitive for a \emph{single}
switch. \Lang offers a network-wide OBS programming model, with
a compiler to distribute the programs across the network. Thus,
although \Lang is exponentially more compact than a state machine
in cases where state is indexed by contents of packet header fields, both FAST
and OpenState can be used as a target for a subset of \Lang programs.

\snaptitle{Optimizing placement and routing.} 
Several projects have
explored optimizing placement of middleboxes and/or routing traffic through
them. 
These projects and \lang share the mathematical problem of placement
and routing on a graph. 
Merlin programs specify
service chains as well as optimization objectives \cite{Merlin}, 
and the compiler uses an MILP to 
choose paths for traffic with respect to specification. 
However, it does not decide the placement of service boxes itself.
Rather, it chooses the paths to pass through the existing instances of 
the services in the physical network.
Stratos \cite{Stratos} explores middlebox placement and distributing
flows amongst them to minimize inter-rack traffic, and 
Slick \cite{Slick} breaks middleboxes into
fine-grained elements and distributes them across the network while
minimizing congestion.
However, they both have a separate algorithm for placement.
In Stratos, placement results is used in an ILP to decide distribution of flows. 
Slick uses a virtual topology on the placed elements with heuristic link weights, 
and finds shortest paths between traffic endpoints.
\section{Conclusion}\label{sec:conclusion}

In this paper, we introduced a stateful SDN programming model with a one-big-switch abstraction, persistent global arrays, and network transactions.  We developed algorithms for analyzing and compiling programs, and distributing their state across the network.  Based on these ideas, we prototyped and evaluated the SNAP language and compiler on numerous sample programs. \cam{We also explore several possible extensions to \lang to support a wider range of stateful applications. Each of these extensions introduces new and interesting research problems to extend our language, compilation algorithms, and prototype.}

\section*{Acknowledgments}

This work was supported by NSF CNS-1111520 and gifts from Huawei, Intel, and Cisco. We thank our SIGCOMM'16 shepherd, Sujata Banerjee, and the anonymous SIGCOMM'16 reviewers for their thoughtful feedback; Changhoon Kim, Nick \nohyphens{McKeown}, Arjun Guha, and Anirudh Sivaraman for helpful discussions; and Nick Feamster, Ronaldo \nohyphens{Ferreira}, Srinivas Narayana, and Jennifer Gossels for feedback on earlier drafts.
\label{lastpage}

\end{sloppypar}

\setlength{\parskip}{-1pt}
\scriptsize
\bibliography{paper}
\bibliographystyle{abbrv}

\ifappendix
\clearpage
\onecolumn
\appendix
\section{Formal Semantics of \Lang}
\label{app:semantics}

\begin{figure}[ht!]
\begin{mdframed}
\scriptsize
\centering
\begin{minipage}{\textwidth}
\[\begin{array}{rcl}
  v \in \Val &::=& 
    \textsf{IP addresses} \, | \, 
    \textsf{TCP ports} \, | \, 
    \dots \, | \,
    \harpoon{v} \\
  l \in \Log & ::= & \elog \, | \, R \, s \cup l \, | \, W \, s \cup l \\
  && \\
  E \cup l &=& l \\
  (R \, s , l_1) \cup l_2 &=& l_1 \cup (R \, s, l_2) \\
  (W \, s , l_1) \cup l_2 &=& l_1 \cup (W \, s, l_2)
\end{array} \]
\end{minipage}
\,
\fbox{$\evale : \Expr \rightarrow \Packet \rightarrow \Val$}
\[\begin{array}{rcl}
  \evale(v,pkt) &=& v \\
  \evale(f,pkt) &=& pkt.f \\
  \evale(\harpoon{e},pkt) &=& \evale(e_1,pkt),\dots,\evale(e_n,pkt) \\
  \multicolumn{3}{r}{\text{where } \harpoon{e} = e_1,\dots,e_n}
\end{array}\]

\fbox{$\eval : \Pol \rightarrow \Store \rightarrow \Packet \rightarrow \mathsf 
               \Store \times 2^{\Packet} \times \Log$}
\[\begin{array}{lll}

\eval(0, m, pkt) & = & (m, \emptyset, \elog) \\
\eval(1, m, pkt) & = & (m, \set{pkt}, \elog) \\

\eval(f = v, m, pkt) & = & 
  (m, 
   \begin{cases}
     \set{pkt} & pkt.f = v \\
     \emptyset & \text{otherwise}
   \end{cases},
   \elog) \\

\eval(s[e_1] = e_2, m, pkt) & = & 
  (m, 
   \begin{cases}
      \set{pkt} & m ~ s ~ \evale(e_1, pkt) = \evale(e_2, pkt) \\
      \emptyset & \text{otherwise}
   \end{cases}, 
   R\, s) \\

\eval(\neg a, m, pkt) & = & 
  \text{ let } (\_, PKT, l) = \eval(a, m, pkt) \text{ in } 
  (m, \set{pkt} \setminus PKT, l) \\ 

\eval(f \leftarrow v, m, pkt) & = & (m, pkt[f \mapsto v], \elog) \\

\eval(s[e_1] \leftarrow e_2, m, pkt) & = & 
  (\lambda s'. \lambda e'. 
   \begin{cases}
     \evale(e_2,pkt) & s = s' \land e' = \evale(e_1, pkt) \\
     m ~ s' ~ e' & \text{otherwise} 
   \end{cases}, 
   \set{pkt}, 
   W\, s) \\\\

\eval(s[e_1]\pp, m, pkt) & = & 
  (\lambda s'. \lambda e'. 
   \begin{cases}
     (m ~ s' ~ e') + 1& s = s' \land e' = \evale(e_1, pkt) \\
     m ~ s' ~ e' & \text{otherwise} 
   \end{cases}, 
   \set{pkt}, 
   W\, s) \\\\

\eval(s[e_1]\mm, m, pkt) & = & 
  (\lambda s'. \lambda e'. 
   \begin{cases}
     (m ~ s' ~ e') - 1 & s = s' \land e' = \evale(e_1, pkt) \\
     m ~ s' ~ e' & \text{otherwise} 
   \end{cases}, 
   \set{pkt}, 
   W\, s) \\

\eval(\IfElse{a}{p}{q}, m, pkt) & = & 
      \text{let } (m',  PKT,  l)  = \eval(a, m, pkt) \text{ in } \\
  & & \text{let } (m'', PKT', l') = \begin{cases}
        \eval(p, m', pkt) & PKT = \set{pkt} \\
        \eval(q, m', pkt) & PKT = \emptyset \end{cases} \\
  & & \text{in } (m'', PKT', l' \cup l) \\\\

\consistent(l_1, l_2) & = & 
      \forall s, (W\, s \in l_1 \implies (R\, s \notin l_2 \land W\, s \notin l_2)) \\
  & & \land (W\, s \in l_2 \implies (R\, s \notin l_1 \land W\, s \notin l_1)) \\

\merge(m, m_1, m_2) & = & 
  \lambda s. \begin{cases}
    m_2 \, s & \forall e, \, m_1 \, s \, e = m \, s \, e \\
    m_1 \, s & \text{otherwise}
  \end{cases} \\
\merge(m, m_1, m_2, \dots, m_k) & = & \merge(m, m_1, \merge(m, m_2, \dots, m_k)) \\

\eval(p + q, m, pkt) & = & 
     \text{let } (m_1, PKT_1, l_1) = \eval(p, m, pkt) \text{ in } \\
  && \text{let } (m_2, PKT_2, l_2) = \eval(q, m, pkt) \text{ in }\\
  && \qquad \begin{cases}
       (\merge(m, m_1, m_2), PKT_1 \cup PKT_2, l_1 \cup l_2) & \consistent(l_1, l_2) \\
       \bot & \text{otherwise}
     \end{cases} \\

\eval(p ; q, m, pkt) 
  & = & \text{let } (m_1, PKT_1, l_1) = \eval(p, m, pkt) \text{ in } \\
  && \text{let } (m_{21}, PKT_{21}, l_{21}),\dots,(m_{2n}, PKT_{2n}, l_{2n}) = \\
  && \eval(q, m_1, pkt_1 \in PKT_1),\dots,\eval(q, m_1, pkt_n \in PKT_1) \text{ in } \\
  && \qquad \begin{cases}
    (\merge(m,m_{21},\dots,m_{2n}), \bigcup_{i=1}^n PKT_{2i}, l_1 \cup (\bigcup_{i=1}^n l_{2i}))
      & \forall i \ne j, ~ \consistent(l_{2i},l_{2j}) \\
    \bot & \text{otherwise}
  \end{cases} \\

\eval(\atomic(p), m , pkt) &=& \eval(p, m, pkt)
\end{array}\]
\end{mdframed}
\captionsetup{skip = 3mm}
\caption{\Lang Semantics}
\label{fig:semantics}

\end{figure} 

\twocolumn

\clearpage
\section{State Dependency Algorithm}\label{app:st-dep}
\normalsize
\begin{figure}[h!]
\scriptsize
\mdfsetup{
innerleftmargin=0mm,
skipabove=0mm,
innertopmargin=-2mm,
skipbelow = 3mm,
}
\begin{mdframed}
\[\begin{array}{rcl} 

\stgraph(\union{p}{q}) & = & \stgraph(p) \cup \stgraph(q) \\
\stgraph(\seq{p}{q}) 
   & = & (\R(p) \times \W(q)) \cup {} \\
      && \stgraph(p) \cup \stgraph(q) \\
\multicolumn{3}{l}{\stgraph(\IfElse{a}{p}{q}) = {(\R(a) \times (\W(p) \cup \W(q)))}} \\

& &	\cup~\stgraph(p) \cup \stgraph(q) \\
\stgraph(\atomic(p)) 
  &=& (\R(a) \cup \W(a)) \times (\R(a) \cup \W(a)) \\
\stgraph(p) & = & \emptyset \text{ otherwise}\\[0.5em]
\R(p) & : & \text{set of state variables read by $p$}\\
\W(p) & : & \text{set of state variables written by $p$}
\end{array}\]
\end{mdframed}
\caption{\stgraph\ function for determining ordering constraints against state variables.}
\label{fig:st-graph}
\end{figure}

\section{Extended State Sharding}\label{app:sharding}
Consider $s[inport]$ for instance. The compiler 
partitions $s$ into $s_1$ to $s_k$, where $s_i$ stores
$s$ for port $i$. The MILP can be used as before to decide 
placement and routing, this time with
the option of placing $s_i$'s at different places without worrying about synchronization as
$s_i$s store \emph{disjoint} parts of $s$. 
The same idea can be used
for distributing $t[srcip]$, where $t_1$ to $t_k$ are $t$'s partitions for 
disjoint subset of IP addresses $ip_1$ to $ip_k$. In this case, 
each port $u$ in the OBS should be replaced with $u_1$ to $u_k$, with $u_i$ 
handling $u$'s traffic with source IP $ip_i$.

\section{Deciding Egress Ports}\label{appendix}

One might worry that the it isn't always possible to decide the egress
port $v$ for a given packet upon entry because it depends on the
state. 
Suppose a packet arrives at port 1 in our example topology
and the user policy specifies that its outport should be assigned to
either 5 or 6 based on state variable $s$, located at $C_6$.
Assume the MILP assigns the path $p_1$ to $(1, 5)$ traffic and the
path $p_2$ to $(1, 6)$.  The ingress switch ($I_1$) can not determine
whether the packet belongs to $(1, 5)$ or $(1, 6)$ to forward it on
$p_1$ or $p_2$ respectively. But: it does not actually matter! Both
paths go through $C_6$ because both kinds of traffic need $s$.
In order to ensure better usage of resources, we can choose which of
$p_1$ and $p_2$ to send the packet over in proportion to each path's
capacity. But whichever path we take, the packet will make its way to
$C_6$ and its processing continues from there.

More formally, the MILP outputs the 
optimized path for the traffic between each ingress port $u$ and egress port $v$. However, the policy
may not be able to determine $v$ at the ingress switch. Suppose that 
 $v_1, \cdots, v_k$ are the possible
outport for packets that enter from $u$. From packet-state mapping (section~\ref{sec:psm}),
we know that the packets from $u$ to each $v_i$ need a sequence (as they are now ordered) of state variables 
$\sequence{s_{i1}, \cdots, s_{ip}}$. 
Therefore, the designated path for this traffic goes through the sequence of 
nodes $\sequence{u, n_{i1}, \cdots, n_{ip}, v_i}$ where $n_{ij}$ is the switch holding $s_{ij}$. 
Now suppose that the policy starts processing a packet from inport $u$ and gets stuck on 
a statement containing $s$. If $s$ only appears in $v_i$'s state sequence, the policy's 
getting stuck on $s$ implies that the packet belongs to the traffic from $u$ to $v_i$, so we 
can safely forward the packet 

on its designated path. However, it may be the case that 
$s$ appears in the state sequences of multiple $v_i$s, each at index $l_i$. 
Thus, we have multiple paths to the switch holding $s$, where the path assigned to $(u, v_i)$'s traffic
is $\sequence{u, n_{i1}, \cdots, n_{il_i}}$ and is capable of carrying at least $d_{uv_i}$
volume of traffic. Let's call the set of $v_i$s whose traffic need $s$, $V_s$. The observation here is that at most 
$\sum_{v_i \in V_s} d_{uv_i}$ worth of traffic entering from $u$ needs state $s$, and the 
total capacity of the designated paths from $u$ to $n_{il_i}$, where $s$ is held, is also equal to 
$\sum_{v_i \in V_s} d_{uv_i}$. Therefore, we just send the traffic that needs $s$ over
one of these paths in proportion to their capacity. The packet will make its way to the switch
holding $s$, and its processing will continue from there. A similar technique is used whenever a switch
gets stuck on the processing of a packet because of a state variable that is not locally available. 

\clearpage
\onecolumn
\section{FDD Sequential Composition}
\label{app:fdd-seq}

Figure~\ref{alg:fdd_seq} contains a high-level pseudocode for the base case of sequential composition, namely when composing one action sequence with another FDD. Apart from the composition operands, function \textproc{seq} has a third argument, $T$, which we call \emph{context}. Context is basically a set of pairs, where each pair consists of a test and its result ($y$ for yes if the tests holds, and $n$ for no). While recursively composing the action sequence with the FDD, we accumulate the resulting tests and their results in $T$ to further use them, deeper in the recursion, to find out whether two fields are equal or not, or whether a field is equal to a specific value or not. 

\textproc{seq} uses several helper functions, the pseudocode of many of which are included in this section. We have excluded the details of some helper functions for simplicity. More specifically, \textproc{update} takes a context and a mapping from field to values, and updates the context according to the mapping. For instance, if $f$ is mapped to $v$ in the input mapping, the input context will be updated to include $(f = v, y)$. \textproc{infer} takes a context, a test, and a test result ($y$ or $n$), and returns true if the specified test result can be inferred from the context for the given test. \textproc{value} takes in a context and a field $f$. If it can be inferred from the context that $f = v$, \textproc{value} returns $v$, and returns $f$ otherwise. Finally, \textproc{reverse} reverses the input list. 

\begin{figure}[h!]
\includegraphics[width=\textwidth]{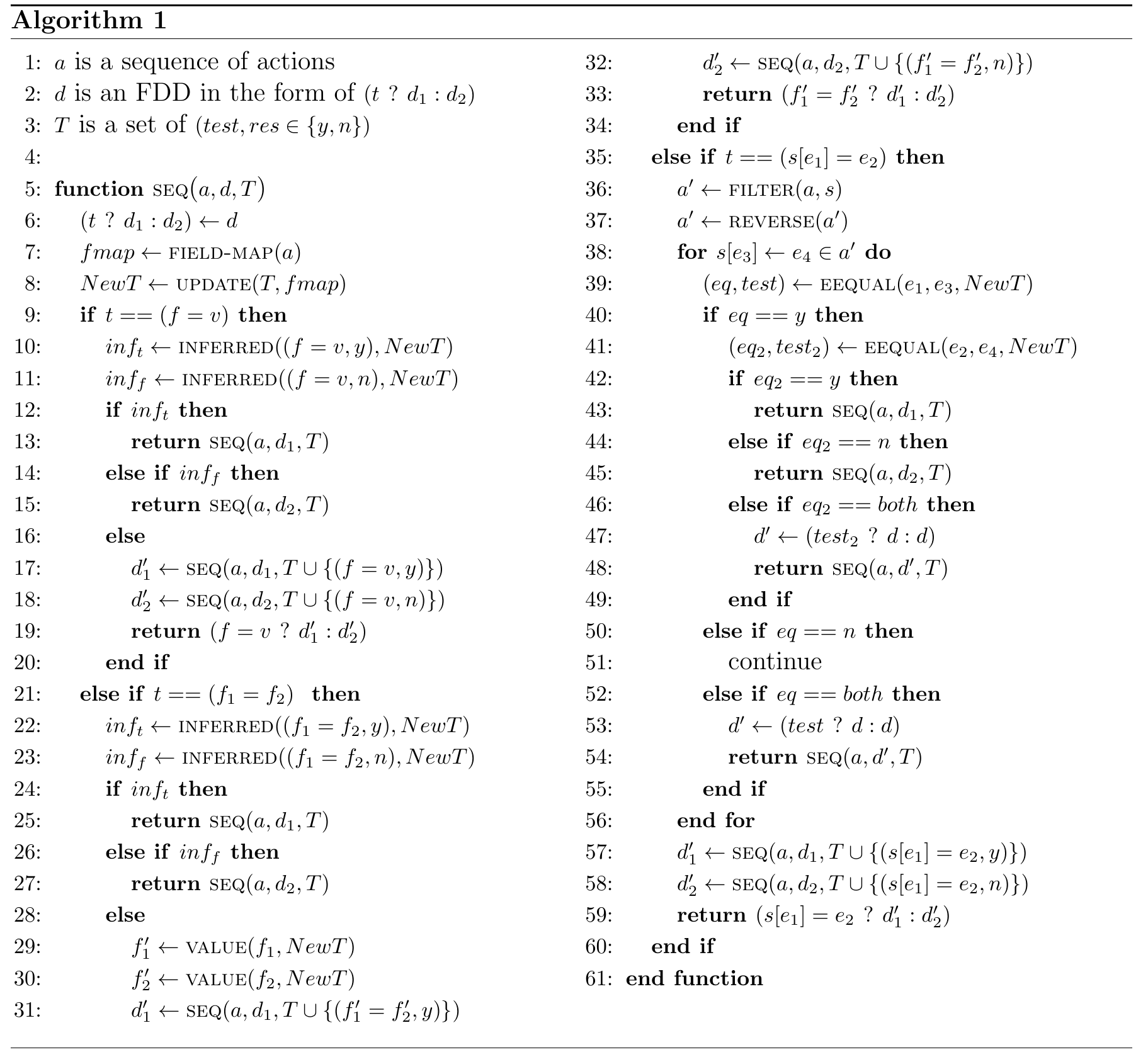}
\caption{Base Case for Sequential Composition of FDDs.\label{alg:fdd_seq}}
\end{figure}

\begin{figure}[h!]
\includegraphics[width=\textwidth]{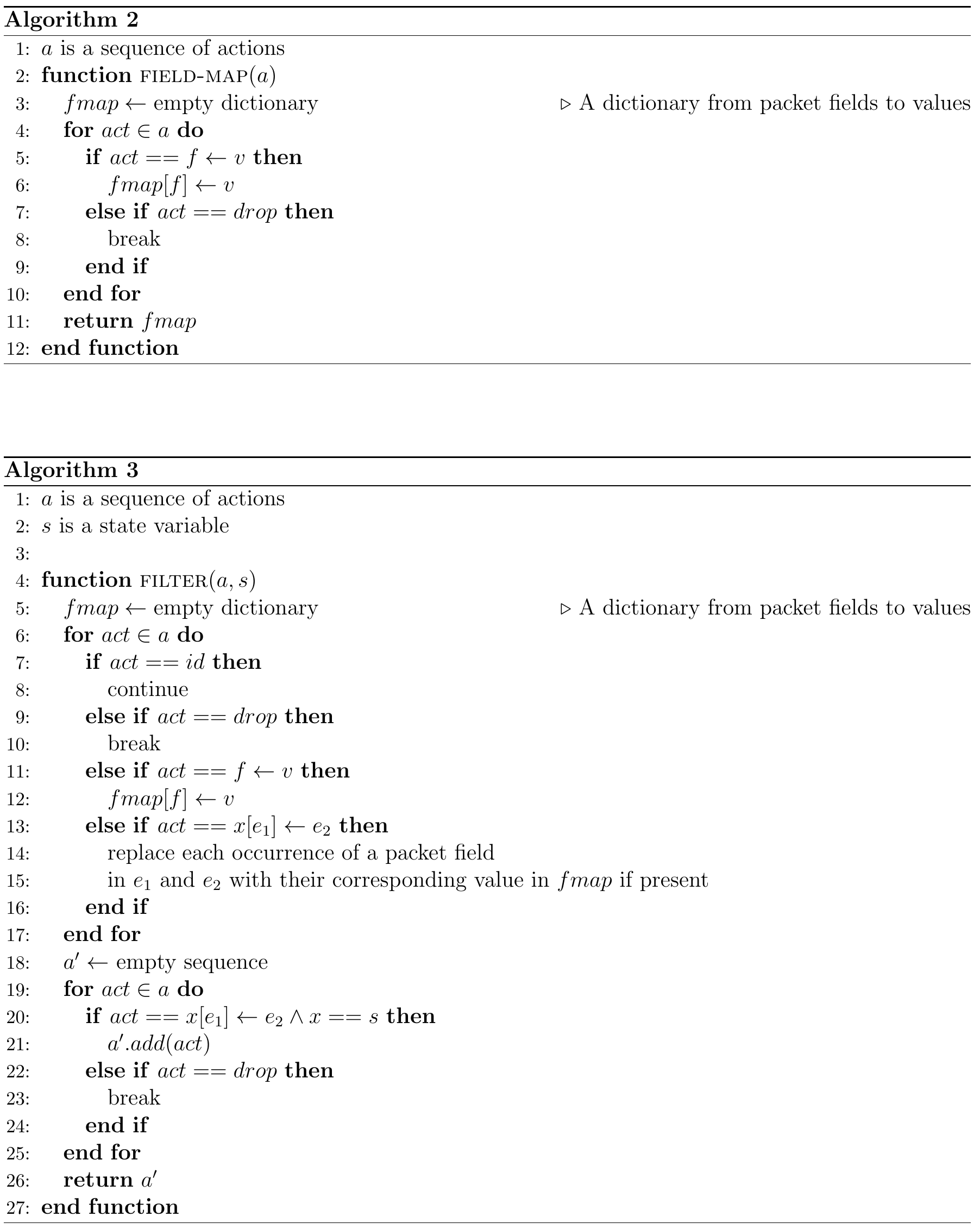}
\end{figure}

\begin{figure}[h!]
\includegraphics[width=\textwidth]{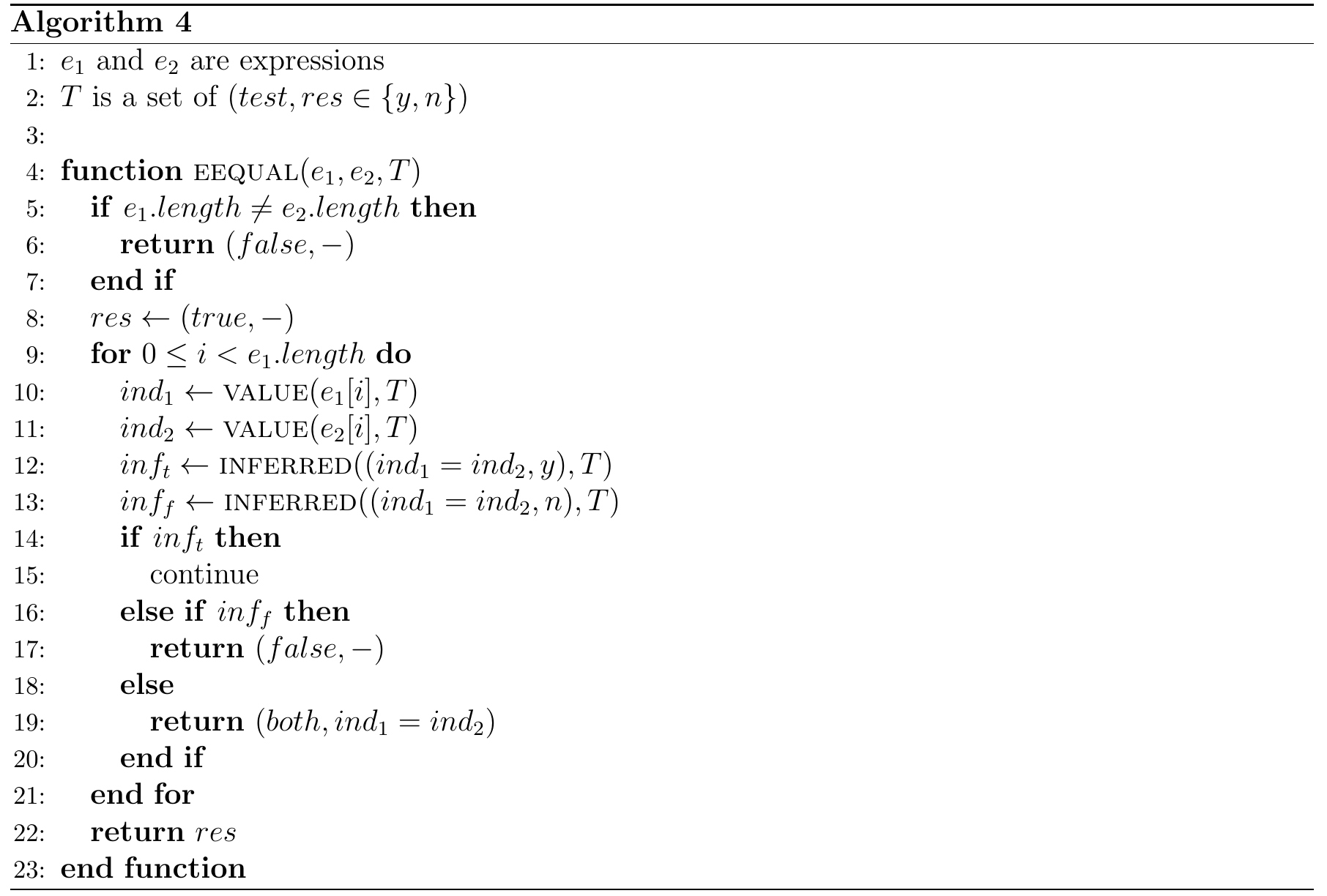}
\end{figure}

\twocolumn

\clearpage
\section{\Lang Policy Examples}
\label{app:examples}

\newenvironment{snappolicy}[1]
{	
\def\savedcaption{\caption{#1}}
\def\savedlabel{\label{alg:#1}}
\begin{algorithm}[h!] 
\begin{algorithmic}[1]%
\scriptsize
}
{\end{algorithmic}
\savedcaption
\savedlabel
\end{algorithm}}

\snaptitle{Number of domains that share the same IP address.}
Suppose an attacker tries to avoid blocking access to his malicious IP through a specific DNS domain by frequently changing the domain name that relates to that IP~\cite{chimera}.
Detection of this behavior is implemented by policy~\ref{alg:many-ip-domains}.
\begin{snappolicy}{many-ip-domains}
	\If{$srcport = 53$}
		\If{$\neg$domain-ip-pair[DNS.rdata][DNS.qname]}
			\State num-of-domains[DNS.rdata]++;
			\State domain-ip-pair[DNS.rdata][DNS.qname] $\gets$ True;
			\If{num-of-domains[DNS.rdata] =threshold}
				\State mal-ip-list[DNS.rdata]$\gets$ True
			\Else \State id
			\EndIf
		\Else \State id
		\EndIf
	\Else  \State id
	\EndIf
\end{snappolicy}

\snaptitle{Number of distinct IP addresses per domain name.}
Too many distinct IPs under the same domain may indiciate a malicious activity~\cite{chimera}. Policy~\ref{alg:many-domain-ips} counts number of different IPs for the same domain name and checks whether it crosses some threshold.

\begin{snappolicy}{many-domain-ips}
	\If{srcport = 53}
		\If{$\neg$ip-domain-pair[DNS.qname][DNS.rdata]}
			\State num-of-ips[DNS.qname]++;
			\State ip-domain-pair[DNS.qname][DNS.rdata] $\gets$ True;
			\If{num-of-ips[DNS.qname] =threshold}
				\State mal-domain-list[DNS.qname]$\gets$ True
			\Else \State id
			\EndIf
		\Else \State id
		\EndIf
	\Else \State id
	\EndIf
\end{snappolicy}

\snaptitle{Stateful firewall.}
A stateful firewall for the CS department, implemented by policy~\ref{alg:stateful-fw} allows only connections initiated within ip$_6$, which is the CS department.

\begin{snappolicy}{stateful-fw}
\If{srcip=ip$_6$}
	\State {established[srcip][dstip] $\gets$ True}
\Else 
	\If{dstip=ip$_6$}
		\State {established[dstip][srcip]}
	\Else \State {id}
	\EndIf
\EndIf
\end{snappolicy}

\snaptitle{DNS TTL change tracking}.  The frequency of TTL changes in the
DNS response for a domain is a feature that can help identify a
malicious domain~\cite{chimera}. Policy~\ref{alg:dns-ttl-change}
keeps track of the number of changes in the announced TTL for each
domain and in the \codeb{ttl-change} state variable. This state
variable can be used in the subsequent parts of the policy to
blacklist a domain.

\begin{snappolicy}{dns-ttl-change}
	\If{srcport = 53}
		\If{$\neg$seen[dns.rdata]}
			\State seen[dns.rdata]$\gets$True;
			\State last-ttl[dns.rdata]$\gets$dns.ttl;
			\State ttl-change[dns.rdata]$\gets$0
		\Else
			\If{last-ttl[dns.rdata] = dns.ttl}
				\State id
			\Else	
				\State last-ttl[dns.rdata]$\gets$ dns.ttl;
				\State ttl-change[dns.domain]++
			\EndIf
		\EndIf	
	\Else \State id
	\EndIf
\end{snappolicy}

\snaptitle{FTP monitoring.}
Policy~\ref{alg:ftp-monitoring} tracks the states of FTP control channel and allows data channel traffic
only if there has been a signal on the control channel. 
The policy assumes FTP standard mode where client announces data port (ftp.PORT), 
other complicated modes may be implemented as well.

\begin{snappolicy}{ftp-monitoring}
	\If{dstport=21}
		\State {ftp-data-chan[srcip][dstip][ftp.PORT]$\gets$True}
	\Else
		\If{srcport=20}
			\State {ftp-data-chan[dstip][srcip,][ftp.PORT]}
		\Else \State {id}
		\EndIf			
	\EndIf
\end{snappolicy}

\snaptitle{Phishing/spam detection.}
To detect suspicious Mail Transfer Agents (MTAs), policy~\ref{alg:spam-detection} detects new MTAs, then checks if any of them sends a large amount of mails in its first 24 hours.
We assume state variables will be reset every 24 hours. 

\begin{snappolicy}{spam-detection}
	\If{MTA-dir[smtp.MTA] = Unknown}
		\State MTA-dir[smtp.MTA] $\gets$ Tracked; 
		\State mail-counter[smtp.MTA] = 0
	\Else \State id;
	\EndIf
	\If{MTA-dir[smtp.MTA] = Tracked}
		\State mail-counter[smtp.MTA]++; 
		\If{mail-couter[smtp.MTA] = threshold}
			\State MTA-directory[smtp.MTA] $\gets$ Spammer
			\Else \State {id}
			\EndIf
	\Else \State id
	\EndIf
\end{snappolicy}

\snaptitle{Heavy hitter detection.}
Policy~\ref{alg:heavy-hitter-detection} keeps a counter per flow and 
marks those passing a threshold as heavy hitters.
To detect and block heavy hitters, one could use the following policy:
\codeb{heavy-hitter-detection; \\ (heavy-hitter[srcip] = False)}

\begin{snappolicy}{heavy-hitter-detection}
	\If{tcp.flags = SYN \& $\neg$heavy-hitter[srcip]}
		\State hh-counter[srcip] ++;
		\If{hh-counter[srcip] = threshold}
			\State {heavy-hitter[srcip] $\gets$ True}
		\Else \State {id}
		\EndIf
	\Else \State {id}
	\EndIf
\end{snappolicy}

\snaptitle{Sidejack detection.} Sidejacking occurs when an attacker
steals the session id information from an unencrypted HTTP cookie and
uses it to impersonate the legitimate user.  Sidejacking can be
detected by keeping track of the client IP address and user agent for
each session id, and checking subsequent packets for that session id
to make sure they are coming from the client that started the
session~\cite{chimera}. This procedure is implemented in \Lang in policy~\ref{alg:sidejacking}.

\begin{snappolicy}{sidejacking}
	\If{(dstip = server) \& $\neg$(sid\footnote{\codeb[scriptsize]{sid = http.session-id}} = null)}
		\If{$\neg$active-session[sid]}
			\State atomic(active-session[sid]$\gets$True;
			\State sid2ip[sid]$\gets$srcip;
			\State sid2agent[sid]$\gets$http.user-agent)
		\Else
			\If{sid2ip[sid] = srcip \& sid2agent[sid] = http.user-agent}
			\State {id}
			\Else \State {drop}
			\EndIf
		\EndIf
	\Else \State {id}
	\EndIf
\end{snappolicy}

\snaptitle{Super-spreader detection.}
Policy~\ref{alg:super-spreader-detection} increases a counter on SYNs and decreases it on FINs per IP address that initiates the connection. 
If an IP address creates too many connections without closing them, it is marked as a super spreader.

\begin{snappolicy}{super-spreader-detection}
	\If{tcp.flags=SYN}
		\State spreader[srcip]++;
		\If{spreader[srcip] = threshold}
			\State {super-spreader[srcip] $\gets$ True}
		\Else \State {id}
		\EndIf
	\Else 
		\If{tcp.flags=FIN}
			\State {spreader[srcip]-{}-}
		\Else \State {id}
		\EndIf
	\EndIf
\end{snappolicy}
\snaptitle{Sampling based on flow-size.}
Policy~\ref{alg:sampling-based-flow-size} uses policy~\ref{alg:flow-size-detect} to detect flow size by keeping a counter for a flow size and select sampling rate based on the counter value. Then it uses one of the three sampler policies~\ref{alg:sample-small}
,~\ref{alg:sample-medium}, or ~\ref{alg:sample-large} for differentiated sampling.

\begin{snappolicy}{flow-size-detect}
	\State flow-size[flow-ind\footnotemark]++;
	\If{flow-size[flow-ind]=1}
		\State {flow-type[flow-ind]$\gets$SMALL}
	\Else \If{flow-size[flow-ind]=100}
		\State {flow-type[flow-ind]$\gets$MEDIUM}
		\Else \If{flow-size[flow-ind]=1000}
			\State {flow-type[flow-ind]$\gets$LARGE}
			\Else \State {id}
			\EndIf
		\EndIf
	\EndIf
\end{snappolicy}
\begin{snappolicy}{sampling-based-flow-size}
	\State flow-size-detect;
	\If{flow-type[flow-ind\footnotemark[\value{footnote}]]=SMALL}
		\State {sample-small}
	\Else \If{flow-type[flow-ind]=MEDIUM}
		\State {sample-medium}
		\Else \State {sample-large}
		\EndIf
	\EndIf
\end{snappolicy}

\begin{snappolicy}{sample-small}
	\State small-sampler[flow-ind\footnotemark[\value{footnote}]]++;
	\If{small-sampler[flow-ind]=5}
		\State {small-sampler[flow-ind]$\gets$0}
	\Else \State {drop}
	\EndIf
\end{snappolicy}

\begin{snappolicy}{sample-medium}
	\State medium-sampler[flow-ind\footnotemark[\value{footnote}]]++;
	\If{medium-sampler[flow-ind]=50}
		\State {medium-sampler[flow-ind]$\gets$0}
	\Else \State {drop}
	\EndIf
\end{snappolicy}

\begin{snappolicy}{sample-large}
	\State large-sampler[flow-ind\footnotemark[\value{footnote}]]++;
	\If{large-sampler[flow-ind]=500}
		\State {large-sampler[flow-ind]$\gets$0}
	\Else \State {drop}
	\EndIf
\end{snappolicy}
\footnotetext{[flow-ind] = [srcip][dstip][srcport][dstport][port]}
\snaptitle{Selective packet dropping.}
Policy~\ref{alg:selective-packet-dropping} drops differentially-encoded B frames in an MPEG encoded stream if the dependency (preceding I frame) was dropped.

\begin{snappolicy}{selective-packet-dropping}
	\If{mpeg.frame-type=Iframe}
		\State {dep-count[srcip][dstip][srcport][dstport]$\gets$14}
	\Else \If{dep-count[srcip][dstip][srcport][dstport]=0}
			\State {drop}
		\Else \State {dep-count[srcip][dstip][srcport][dstport]-{}-}
		\EndIf
	\EndIf
\end{snappolicy}
\snaptitle{Connection Affinity.}
Policy~\ref{alg:conn-affinity} uses TCP state machine to distinguish ongoing connections from new ones, assuming \codeb{basic-tcp-reassembly ; conn-affinity} and
that we want to do per-connection load balancing using \codeb{lb}.

\begin{snappolicy}{conn-affinity}
	\If{tcp-state[dstip][srcip][dstport][srcport][proto] = ESTABLISHED $|$ tcp-state[srcip][dstip][srcport][dstport][proto] = ESTABLISHED}
		\State {lb}
	\Else \State {id}
	\EndIf
\end{snappolicy}
\snaptitle{SYN flood detection.}
To detect SYN floods, we should count the number of SYNs without any matching ACK from the sender side and if this sender crosses a certain threshold it should be blocked.
This can actually be implemented in a similar way as the super-spreader-detection policy (policy~\ref{alg:super-spreader-detection}).

\snaptitle{Elephant flow detection.}
Suppose the attacker launches legitimate
but very large flows. One could detect abnormally
large flows, flag them as attack flows, and then
randomly drop packets from these
large flows.
This policy can actually be implemented by a composition of previously implemented
policies: \\ \codeb{flow-size-detect;sample-large policy}.

\snaptitle{DNS amplification mitigation}
In a DNS amplification attack, the attacker spoofs and sends out many DNS queries with the IP address of the victim. Thus, large answers are
sent back to the victim that can lead to denial of service in case the victim is a server.
Policy~\ref{alg:dns-amplification} detects this attack by tracking the DNS queries that the server has actually sent out, and getting suspicious of attack
if the number of unmatched DNS responses passes a threshold.

\begin{snappolicy}{dns-amplification}
	\If{dstport=53}
		\State {benign-request[srcip][dstip]$\gets$True}
	\Else \If{srcport=53 \& $\neg$benign-request[dstip][srcip]}
			\State {drop}
		\Else \State {id}
		\EndIf
	\EndIf
\end{snappolicy}

\snaptitle{UDP flood mitigation.}
Policy~\ref{alg:udp-flood} identifies
source IPs that send an anomalously higher number
of UDP packets and uses this to categorize each
packet as either attack or benign.

\begin{snappolicy}{udp-flood}
	\If{proto = UDP \& $\neg$udp-flooder[srcip]}
		\State udp-counter[srcip] ++;
		\If{udp-counter[srcip] = threshold}
			\State udp-flooder[srcip] $\gets$ True;
			\State drop
		\Else \State {id}
		\EndIf
	\Else \State {id}
	\EndIf
\end{snappolicy}
\snaptitle{Snort flowbits.}
The Snort IPS rules~\cite{Snort}
contain both stateless and stateful ones. Snort uses a tag called \emph{flowbits} to mark a boolean state of a ``5-tuple''. 
The following example shows how flowbits are used for application specification: 

\mdfsetup{
skipabove = 3mm,
skipbelow = 3mm,
}
\begin{mdframed}
	\centering
	\codeb[scriptsize]{pass tcp HOME-NET any -> EXTERNAL-NET 80 \\
		(flow:established; content:"Kindle/3.0+"; \\ \textbf{flowbits:set,kindle};) \\} 
\end{mdframed}

The same rule can be expressed in \Lang terms as can be seen in policy~\ref{alg:snort-flowbits}:

\begin{snappolicy}{snort-flowbits}
\State \match{srcip}{HOME-NET};
\State \match{dstip}{EXTERNAL-NET};
\State \match{dstport}{80}; 
\State \match{established[srcip][dstip][srcport][dstport][proto]}{True};
\State \match{content}{"Kindle/3.0+"} ;
\State \textbf{kindle[srcip][dstip][srcport][dstport][proto]} $\gets$ \textbf{True}
\end{snappolicy}
Note that Snort's flowbits are more restricted than \Lang state variables in the sense that  they can only be defined per 5-tuple, i.e.\ the index to the state is fixed. 

\newpage
\snaptitle{Basic TCP state machine.}
Policy~\ref{alg:basic-tcp-reassembly} implements a basic bump-on-the-wire TCP state machine.
\onecolumn

\begin{snappolicy}{basic-tcp-reassembly}
	\If{tcp.flags=SYN \&  tcp-state[srcip][dstip][srcport][dstport][proto] =CLOSED}
		\State {tcp-state[srcip][dstip][srcport][dstport][proto] $\gets$ SYN-SENT}
	\Else \If{tcp.flags=SYN-ACK \& 		
			tcp-state[dstip][srcip][dstport][srcport][proto]=SYN-SENT}
		\State {tcp-state[dstip][srcip][dstport][srcport][proto]$\gets$SYN-RECEIVED}
	\Else \If{tcp.flags=ACK \&	
				tcp-state[srcip][dstip][srcport][dstport][proto]=SYN-RECEIVED}
		\State {tcp-state[srcip][dstip][srcport][dstport][proto]$\gets$ESTABLISHED}
	\Else \If{tcp.flags=FIN \& 		
					tcp-state[srcip][dstip][srcport][dstport][proto]=ESTABLISHED}
		\State {tcp-state[srcip][dstip][srcport][dstport][proto]$\gets$FIN-WAIT}
	\Else \If{tcp.flags=FIN-ACK \& 		
						tcp-state[dstip][srcip][dstport][srcport][proto]=FIN-WAIT}
		\State {tcp-state[dstip][srcip][dstport][srcport][proto]$\gets$FIN-WAIT2}
	\Else \If{tcp.flags=ACK \& 		
							tcp-state[srcip][dstip][srcport][dstport][proto]=FIN-WAIT2}
		\State {tcp-state[srcip][dstip][srcport][dstport][proto]$\gets$CLOSED}
	\Else \If{tcp.flags=RST \& 		
								tcp-state[dstip][srcip][dstport][srcport][proto]=ESTABLISHED}
		\State {tcp-state[dstip][srcip][dstport][srcport][proto]$\gets$CLOSED}
		\Else \State {tcp-state[dstip][srcip][dstport][srcport][proto]=ESTABLISHED + tcp-state[srcip][dstip][srcport][dstport][proto]=ESTABLISHED}
		\EndIf \EndIf \EndIf \EndIf \EndIf \EndIf \EndIf					
\end{snappolicy}

\fi
\end{document}